\documentclass[aps,pra,reprint,showpacs,superscriptaddress]{revtex4-1}

\usepackage{amsmath}
\usepackage{amssymb}
\usepackage{bm}
\usepackage{setspace}
\usepackage{amsfonts}
\usepackage{rotating}
\usepackage{amsthm}
\usepackage[english]{babel}

\renewcommand{\vec}[1]{\bm{#1}}
\usepackage{graphicx}
\usepackage{epstopdf}

\begin{document}

\title{Pauli equations and non-commutative position operators in 2D Dirac-like 
semiconductors in view of second quantization}
\author{E.~L.~Rumyantsev}
\author{P.~E.~Kunavin}
\affiliation{Institute of Natural Sciences, Ural Federal University, 620000
Ekaterinburg, Russia}

\begin{abstract}
The Pauli equations describing electron (hole) dynamics in 2D Dirac-like 
intrinsic semiconductors in external (impurity) scalar potential and for 
inhomogeneous lattice distortions are obtained within second quantization 
approach. We show that the modifying external perturbation terms in formulated 
no-pair equations are in general non-local and demonstrates singular behavior in 
gapless situation where they do not depend on semiconductor parameters. It is 
shown that lattice distortion perturbation can cause confinement of both 
electrons and holes in the same spatial region. The proposed approach is 
verified by comparison with well-established results in low energy limit. The 
consideration of position operator in second quantization approach allows 
elucidating the physical meaning of spin-orbit-like and Darwin-like terms.
\end{abstract}


\maketitle

\section{Introduction}

The Pauli single particle equations (SPE) emerge in atomic physics as a 
non-relativistic approximation to the ``exact'' relativistic Dirac equation. 
The truncated $2 \times 2$ Hamiltonian describes only electrons or positrons 
(holes). In obtained Schrodinger-like equation the external potential (in atomic 
physics the potential of nucleus) is modified. The famous term known as 
spin-orbit interaction (SOI):  
$\frac{\hbar}{4m_0^2c^2}\vec{\sigma}\left[\vec{\triangledown}
V(r)\times \hat{\vec{p}}\right]$ is added to “bare” potential. The inclusion of 
this term is necessary for correspondence of theoretical results with 
experimental data \cite{Winkler_SO_Coupl_eff}. The problem of reduction of 
multicomponent relativistic problem constructing approximate quasi-relativistic 
two component Hamiltonian has been and is under intensive discussion in low and 
intermediate nuclear physics and in molecule physics for heavy-atom species. It 
is the long established fact that ``quasi-relativistic'' problems infest any 
multiband semiconductor $\vec{k} \cdot \vec{p}$ Hamiltonians 
\cite{Katsnelson20073,Zawadzki2011}. The band structure of semiconductors 
described in $\vec{k} \cdot \vec{p}$ theory by matrix Hamiltonians takes into 
account at least two (electron and hole-like) bands. Such Dirac-like artifacts 
as constant carrier velocity and Zitterbewegung emerge if we proceed along the 
``classical'' line of approach, allowing for interference of positive and 
negative states \cite{Zawadzki2011_ZB}. The Klein paradox – unimpeded 
penetration of carriers through high potential barrier, emerges as an essential 
attribute of graphene physics described within gapless Dirac-like  Hamiltonian 
\cite{Kats06}. According to seminal paper by Keldysh \cite{Keldysh1963} the 
problem of supercharged nuclei transforms in semiconductors into the problem of 
deep levels of impurity centers. Thus, the problem of reduction of generic 
multicomponent Hamiltonians, which not only makes the computation feasible but 
elucidates the underlying physics, emerges as the interdisciplinary problem. 

The standard derivation of Pauli SPE in Dirac theory follows two main 
approaches. The first approach is the construction of appropriate unitary 
transformation decoupling positive and negative modes of Dirac equation. The 
main schemes are famous Foldy-Wouthuysen (FW) transformation and its variant 
proposed by Eriksen, Duglas-Kroll-Hess approach 
\cite{Foldy1950,Douglas197489,PhysRevA.33.3742,Kutzelnigg1990,Fearing1994657,
Barysz2001,Silenko2003,PhysRevA.90.012112,Liu201459}. The truncated Hamiltonian 
even in the lowest in $c^{-1}$ order suffers from such drawbacks as lack of 
uniqueness and difficulties in obtaining higher order terms especially, if 
external potential is applied. It has no well-behaved wave functions and only 
expectation values obtained with these functions have physical meaning 
\cite{Kutzelnigg1990}. The analog of this approach is used for constructing 
reduced $\vec{k} \cdot \vec{p}$ Hamiltonians for degenerate spectrum 
\cite{Pikus1974}. The second approach is the Pauli subtraction method (small 
component method) which eliminates two components, considered to be small, from 
the four-component wave function \cite{Pauli1958,Rel_Quant_Theory}. Within this 
method block-diagonal energy-dependent Hamiltonians are constructed. Under this 
transformation non-Hermitian terms may appear \cite{PhysRevA.90.012112}. Known 
as Lowdin partitioning \cite{Lowdin1962}, this approach is used usually for 
reduction of multicomponent $\vec{k} \cdot \vec{p}$  Hamiltonians.

The commonly accepted point of view is that Dirac equation is not the single 
particle equation. Rigorously, the Dirac equation is self-consistent only in the 
context of quantum field theory (QFT), especially if external potential is 
applied. This is due to nonzero probability of particle-antiparticle 
creation/annihilation. The space-time resolved solutions to relativistic QFT 
allows to clarify the physics of Klein paradox and Zitterbewegung effect 
\cite{Krekora2004,PhysRevA.59.604}

Bearing in mind discussed above quasi-relativistic similarity of the problems 
arising in  $\vec{k} \cdot \vec{p}$  description, we approach the problem of 
formulation of Pauli electron(hole)-only equations in the theory with external 
perturbations from the point of QFT using the second quantization method (SQM).
The account for filled valence band allows us to use electron-hole semiconductor 
language as the quantum number ``charge'' is ascribed to the eigenfunctions of 
appropriate reduced two-component Pauli Hamiltonians.  As pointed by Feynman 
``The problem of charges in a fixed potential is usually treated by the method 
of second quantization of electron field, using the ideas of the theory of 
holes'' \cite{Feynman1949}.

Note that the second quantization prohibits in general the direct probabilistic 
interpretation of the field operators, as compared to commonly used quantum 
mechanical wave functions \cite{PhysRevA.73.022114}. The superposition of the 
states with variable number of particles is not compatible with the simple 
interpretation of wave function. Moreover the possible interpretation crucially 
depends on the choice of single particle projection space 
\cite{PhysRevA.73.022114}.

In this paper the sought SPE are obtained neglecting pair creation processes 
(stability of vacuum under external perturbation) so they are logically to be 
called as ``no -pair equations''. The decoupling of electrons and holes degrees 
of freedom enables us to write down the Pauli-like equations for each of these 
carriers separately.

The paper is organized as follows. In Sec. \ref{sec:2}, we obtain single 
particle Pauli equations for 2D Dirac-like semiconductor and compare our 
results with well-established ones. In Sec. \ref{sec:3}, we analyze 
properties of the position operator and its connection with spin-orbit-like and 
Darwin-like terms.

\section{Pauli equations in 2D Dirac-like semiconductor}
\label{sec:2}

Leaving aside pure relativistic problems Dirac Hamiltonian is
considered as an example of effective $\vec{k} \cdot \vec{p}$
Hamiltonian, describing two-band semiconductor with symmetrical
conduction and valence bands. The Dirac Hamiltonian has the following
form
\begin{equation}
\label{Dirac_H}
  \hat{H}_D =
   \begin{pmatrix}
    \frac{E_g}{2}  & 0     & \gamma k_z & \gamma k_-  \\
    0    & \frac{E_g}{2}   & \gamma k_+ & -\gamma k_z \\
    \gamma k_z & \gamma k_-  & -\frac{E_g}{2} & 0     \\
    \gamma k_+ & -\gamma k_z & 0    & -\frac{E_g}{2}
   \end{pmatrix} ,
\end{equation}
where  $k_{\pm} = k_x \pm ik_y = ke^{\pm i\varphi(\vec{k})}$, $\gamma$ - 
characteristic velocity, $E_g$ -
energy gap. In relativistic Dirac theory $E_g = 2mc^2$ and $\gamma = c$.  Here
and in the following $\hbar = 1$.

To simplify the following analysis, we consider 2D version of Hamiltonian
\eqref{Dirac_H}. Choosing $k_z = 0$ two groups of
states $(e ~1/2, h ~-1/2)$ and $(e ~-1/2, h ~1/2)$ do not mix. The Hamiltonian
for the former group of these states reads
\begin{equation}
\label{Dirac_H_2D}
  \hat{H}_D = \begin{pmatrix}
              \frac{E_g}{2}  &  \gamma k_-   \\
              \gamma k_+    & -\frac{E_g}{2}
            \end{pmatrix} .
\end{equation}
The Hamiltonian matrix for the second group of states is obtained by replacing
$k_y$ by $-k_y$. In this section we consider only first group of states 
described by Hamiltonian \eqref{Dirac_H_2D}.  The energy eigenvalues of
\eqref{Dirac_H_2D} are
\begin{equation}
\label{eq30}
\varepsilon(k)_{1,2} = \pm\sqrt{\frac{E_g^2}{4} + \gamma^2 k^2}
 \equiv \pm \varepsilon(k) .
\end{equation}
The electrons and holes eigenstates are chosen as
\begin{eqnarray}
\varphi_e(\vec{k}) = \begin{pmatrix}
               \cos\varTheta(\vec{k})  \\
               \sin\varTheta(\vec{k})e^{i\varphi(\vec{k})}
              \end{pmatrix} , \\
\varphi_h(\vec{k}) = \begin{pmatrix}
               -\sin\varTheta(\vec{k})e^{-i\varphi(\vec{k})} \\
               \cos\varTheta(\vec{k})
              \end{pmatrix},
\end{eqnarray}
where 
\begin{eqnarray}
\cos\varTheta(\vec{k}) = \frac{\sqrt{\varepsilon(k) + E_g/2}}
                              {\sqrt{2\varepsilon(k)}}, \\
\sin\varTheta(\vec{k}) = \frac{\sqrt{\varepsilon(k) - E_g/2}}
                              {\sqrt{2\varepsilon(k)}}.
\end{eqnarray}

In the special case $E_g = 0$ Hamiltonian \eqref{Dirac_H_2D} is used for the
description of electrons in $K$ valley in graphene
\begin{equation}
\label{Graphene_H}
  \hat{H}_K = \gamma
            \begin{pmatrix}
               0   & k_-   \\
               k_+ & 0
            \end{pmatrix} .
\end{equation}
In this case we are dealing with pseudospin, which is a formal way of taking 
into account the two carbon atoms per unit cell \cite{RevModPhys.81.109}, and 
the quasi-momentum $k$ is measured from Dirac points.
In the second quantization picture the Hamiltonian \eqref{Dirac_H_2D}
for intrinsic semiconductor is
\begin{equation}
  \hat{H}_D = \int \varepsilon(k)\hat{a}^+(\vec{k})\hat{a}(\vec{k}) 
   \frac{d\vec{k}}{(2\pi)^2}
          + \int \varepsilon(k)\hat{b}^+(\vec{k})\hat{b}(\vec{k}) 
                \frac{d\vec{k}}{(2\pi)^2} \  .
\end{equation}
Here $\hat{a}^+(\vec{k}), ~\hat{a}(\vec{k})$ are creation/annihilation
operators for electrons and $\hat{b}^+(\vec{k}), ~\hat{b}(\vec{k})$ are
corresponding operators for holes. Inserting the potential $V(\vec{r})$ into
 “empty” Hamiltonian diagonal we obtain the following additional terms in the 
Hamiltonian
\begin{eqnarray}
\label{Dirac_H_int}
\nonumber
  \hat{H}_{i} &=&
   \int\int \varphi_e^*(\vec{k})\varphi_e(\vec{q}) V(\vec{k} - \vec{q})
     \hat{a}^+(\vec{k}) \hat{a}(\vec{q})\frac{d\vec{k} d\vec{q}}{(2\pi)^4}\\
    \nonumber
    &-&
    \int\int \varphi_h^*(\vec{k})\varphi_h(\vec{q}) V(\vec{k} - \vec{q})
     \hat{b}^+(\vec{-q}) \hat{b}(\vec{-k})\frac{d\vec{k} d\vec{q}}{(2\pi)^4}\\
    \nonumber
     &+&
    \int\int \varphi_e^*(\vec{k})\varphi_h(\vec{q}) V(\vec{k} - \vec{q})
     \hat{a}^+(\vec{k}) \hat{b}^+(\vec{-q})\frac{d\vec{k} d\vec{q}}{(2\pi)^4}\\
      &+&
      \nonumber
    \int\int \varphi_h^*(\vec{k})\varphi_e(\vec{q}) V(\vec{k} - \vec{q})
     \hat{b}(\vec{-k})\hat{a}(\vec{q}) \frac{d\vec{k} d\vec{q}}{(2\pi)^4} . \\
\end{eqnarray}
The terms containing $\hat{a}^+(\vec{k}) \hat{a}(\vec{q})$
and $\hat{b}^+(\vec{q}) \hat{b}(\vec{k})$ describe the processes of scattering
electrons/holes by the potential modified by the presence of filled valence
band. The terms containing $\hat{a}^+(\vec{k}) \hat{b}^+(\vec{-q})$ and
$\hat{b}(\vec{-k}) \hat{a}(\vec{q})$ describe the perturbation of vacuum
(filled valence band) As it seen the second quantization version of 
considered Dirac Hamiltonian is a many-particle Hamiltonian due to possibility 
of electron-hole pairs creation and annihilation by external potential, even 
when inter-particle interaction (as in our work) is not accounted for. Note, 
that under the term ``external'' potential we understood not only external 
action but internal action due to impurities as well.

The explicit expression for modified no-pair electron scattering
potential is
\begin{eqnarray}
V_e&=&\int\int\left(\cos\varTheta(\vec{k})\cos\varTheta(\vec{q})
 + \sin\varTheta(\vec{k})\sin\varTheta(\vec{q})e^{-i\varDelta(\vec{k},\vec{q})}
\right) \nonumber \\
&\times& V(\vec{k}-\vec{q})\hat{a}^+(\vec{k})\hat{a}(\vec{q})
 \frac{d\vec{k}}{(2\pi)^2}\frac{d\vec{q}}{(2\pi)^2}, \label{Dirac_Ve}
\end{eqnarray}
where $\varDelta(\vec{k},\vec{q}) = \varphi(\vec{k}) - \varphi(\vec{q})$. The
expression for holes (positrons) is of the same functional form but has
opposite sign.

Finally, the spectrum for single particle electron problem is 
derived from Heisenberg equation of motion
$$
\frac{d\hat{a}}{dt}=i\left[H_{p},\hat{a}(\vec{k})\right],
$$
where single particle no-pair Pauli Hamiltonian $H_{p}$ for electron runs as 
follows:
\begin{equation}
H_{p}=\int\varepsilon(\vec{k})\hat{a}^+(\vec{k})\hat{a}(\vec{k})
  \frac{d\vec{k}}{(2\pi)^2} + V_e,
\end{equation}
where $\varepsilon(\vec{k})$ and $V_e$ are given by the formulas
\eqref{eq30} and \eqref{Dirac_Ve}, respectively. The solution of
Heisenberg equation is given by the solution of Eigen problem of corresponding 
no-pair Pauli-type equation:
\begin{eqnarray}
&E&\varphi(\vec{k})=\varepsilon(\vec{k})\varphi(\vec{k})\nonumber \\
&+&\int\left(\cos\varTheta(\vec{k})\cos\varTheta(\vec{q})
 + \sin\varTheta(\vec{k})\sin\varTheta(\vec{q})e^{-i\varDelta(\vec{k},\vec{q})}
\right) \nonumber \\
&\times& V(\vec{k}-\vec{q})\varphi(\vec{q})\frac{d\vec{q}}{(2\pi)^2} . 
\label{eq100}
\end{eqnarray}
Modified electron scattering potential \eqref{eq100} loses its translational 
invariance in momentum space and as a corollary becomes non-local in coordinate 
representation. The obtained equation coincides conceptually with the no-pair 
equation  derived in \cite{Baretty1986} for hydrogenic systems within quantum 
electrodynamic (QED), allowing to make rigorous reduction into the system that 
contains a single electron but no positrons. The analogous no-pair equations has 
been derived with free particle projection operators for atomic orbital 
calculations on one-electron atoms \cite{PhysRevA.33.3742}.

As it was pointed out, the Pauli-like equation for holes differs from
\eqref{eq100} by sign. It means that potential perturbation inserted 
into diagonal of $\vec{k} \cdot \vec{p}$ Hamiltonian (even 
when its ``electro-magnetic'' nature is not specialized) discriminates between 
particle charges. Thus, if it is attractive for electrons, it is repulsive for 
holes and vice versa. The account for omitted pair production terms up to the 
second order in $V(k)$ will lead to the appearance of the terms proportional to 
square of potential, which are charge insensitive and act on electrons and holes 
alike. In this approximation the solutions of obtained equation will predict the 
same effects as the equation proposed by L.Keldysh \cite{Keldysh1963} for the 
description of deep levels.

The simultaneous confinement of electrons and holes within proposed approach can 
occur only due to spatial dependence of band gap (mass term in Dirac equation). 
Following \cite{Gosselin2009} we account for smooth inhomogeneous lattice 
distortion by varying band gap: $E_g \rightarrow E_{g0} + \delta E_g(\vec{r})$. 
As a result the following additional terms in no-pair equations for electrons 
and holes appear respectively
\begin{eqnarray}
&\frac{1}{2}&\int\int\left(\cos\varTheta(\vec{k})\cos\varTheta(\vec{q})
 + \sin\varTheta(\vec{k})\sin\varTheta(\vec{q})e^{-i\varDelta(\vec{k},\vec{q})}
\right) \nonumber \\
&\times& \delta E_g(\vec{k}-\vec{q}) \hat{a}^+(\vec{k})\hat{a}(\vec{q})
\frac{d\vec{k}}{(2\pi)^2}\frac{d\vec{q}}{(2\pi)^2},
\end{eqnarray}
\begin{eqnarray}
&\frac{1}{2}&\int\int\left(\cos\varTheta(\vec{k})\cos\varTheta(\vec{q})
 - \sin\varTheta(\vec{k})\sin\varTheta(\vec{q})e^{i\varDelta(\vec{k},\vec{q})}
\right) \nonumber \\
&\times& \delta E_g(\vec{k}-\vec{q}) \hat{b}^+(\vec{k})\hat{b}(\vec{q})
\frac{d\vec{k}}{(2\pi)^2}\frac{d\vec{q}}{(2\pi)^2}.
\end{eqnarray}
Contrary to the statement in \cite{Gosselin2009} this perturbation cannot be 
identified as some electric field. In contrast, the signs for such ``mass'' 
perturbation do not depend on quasi-particle charge and are the same for 
electrons and holes. It means that if confinement occurs for electrons due to 
proposed mechanism, the holes can localize within the same space region. The 
position dependence of band edges and effective masses were originally 
incorporated into $\vec{k} \cdot \vec{p}$ -type analysis for calculation of band 
structure of GaAs-GaAlAs and InAs-GaSb superlattices in \cite{PhysRevB.24.5693}. 
The necessity of the presence of band inversion or mass domain walls for the 
occurrence of localized edge states in graphene was underlined in 
\cite{RevModPhys.82.3045,2016arXiv160304329D,Volkov1985}.

If equation \eqref{eq100} is rewritten in a coordinate
representation, the potential $V_e$ becomes non-local. Thus
instead of obtaining differential Schrodinger-type of equation we are
to consider integro-differential one. Nevertheless, it is possible to hide this
smearing of potential and reduce this equation to the well-known form
used in atomic physics where SOI emerges. This is achieved by the
expansion of $H_{p}$ including the terms up to the second order in
parameter $\gamma k/E_g$. Within this approximation the kinetic part becomes
$\varepsilon(k)\approx E_g/2+\gamma^2 k^2/E_g -\gamma^4 k^4/E_g^3$ and
$V_e$ can be presented as the sum of two terms: ``bare'' potential and
additional term:
\begin{eqnarray}
V_e&\approx&\int\int
V(\vec{k}-\vec{q})\hat{a}^+(\vec{k})\hat{a}(\vec{q})d\vec{k}d\vec{q}-\nonumber\\
&-&\frac{\gamma^2}{2E_g^2}\int\int\left\{2i[\vec{k}\times\vec{q}]_Z-(\vec{k}
-\vec{q})^2\right\}
\nonumber\\ &\times&
V(\vec{k}-\vec{q})\hat{a}^+(\vec{k})\hat{a}(\vec{q})
\frac{d\vec{k}}{(2\pi)^2}\frac{d\vec{q}}{(2\pi)^2}.
\label{eq110}
\end{eqnarray}
The inequality $\gamma k/E_g\ll 1$ has simple physical meaning: the
characteristic ``Compton'' wavelength $\lambda_C=\gamma/E_g$  is small
as compared with the characteristic length of spatial variation of
electron/hole envelope function $\lambda=1/k$. The condition
$\lambda_C\ll \lambda$ is always implied while considering
semiconductors problems within envelope function approximation (EFA).
The corresponding to \eqref{eq110} equation for electron Eigen functions 
(compare with \eqref{eq100}) is a familiar expression with SOI and Darwin terms 
modifying external potential.
\begin{eqnarray}
&&\left(E - \frac{E_g}{2}\right)\varphi(\vec{r})=\left[\frac{\gamma^2p^2}{E_g}-
\frac{\gamma^4p^4}{E_g^3}\right]\varphi(\vec{r})+ \nonumber \\
&+&\left\{V(\vec{r})+\frac{\gamma^2}{E_g^2}[\nabla V(\vec{r})\times\vec{p}]_Z
+\frac{\gamma^2}{2E_g^2}\nabla^2V(\vec{r})\right\}\varphi(\vec{r}) 
\nonumber \\
\label{eq120}
\end{eqnarray}
This correspondence with well known, if trivial result, is presented only for 
validation of consideration proposed which we are going to apply elsewhere to 
more sophisticated $\vec{k} \cdot \vec{p}$ Hamiltonians. 

It is interesting to compare the additional terms induced by scalar potential 
and lattice distortions (spatially dependent band gap) in considered Pauli 
no-pair equations. In order to make our point more clearly let us consider 1D 
variant of considered problem with $k_y = 0$ in low energy range (EFA). In this 
case only Darwin term survives in the expression for modified scalar potential
\begin{equation}
-\int(k_x - q_x)^2 V(k_x - q_x )\hat{a}^+(k_x)\hat{a}(q_x)
   \frac{dk_x}{2\pi}\frac{dq_x}{2\pi}.
\end{equation}
Lattice distortion induces the following term
\begin{equation}
-\int(k_x + q_x)^2 \delta E_g(k_x - q_x )\hat{a}^+(k_x)\hat{a}(q_x)
   \frac{dk_x}{2\pi}\frac{dq_x}{2\pi}.
\end{equation}
The difference in sign has far-reaching consequences. In the corresponding 
Schrodinger-like equation in coordinate space the modifying term for scalar 
potential is $V''_x(x)\varphi(x)$. The band gap variation induces 
additional terms
\begin{equation}
 -\delta E''_g(x)\varphi(x) - 4 \delta E'_g(x)\varphi'(x)
 - 4 \delta E_g(x)\varphi''(x) .
\end{equation}
At first glance, this expression looks very strange if not to remember that 
in $\vec{k} \cdot \vec{p}$ theory the energy gap determines such important 
characteristic as effective mass $m^*$ around conduction (valence) band minimum 
(maximum). In our Dirac problem effective mass is $m^* = E_{g0}/2\gamma^2$. In 
classical theory the kinetic term with spatially dependent mass is 
$-p^2 \delta m(r)/2(m_0^*)^2$. In quantum theory the appearance of $\vec{r}$ 
dependent mass in kinetic term poses the problem of appropriate ordering 
prescription. The best known procedure coping with this arbitrariness in 
Hamiltonian definition has been proposed by Weyl \cite{Lee1981}. For a product 
of operators $\left\{\hat{Q}_1 \ldots  \hat{Q}_n\right\}$ the Weyl ordering is 
defined as the sum over all permutations $P(i_1\ldots i_n)$ of the indices. For 
the product of $p^2$ and $r$ it prescribes the use of the operator $(p^2r + 
2prp +rp^2)/4$. It is easy to verify, that in our approach this problem is 
solved in the favor of of such ordering
\begin{equation}
-\frac{1}{4}\left( \hat{p}^2\frac{\delta m(r)}{2(m_0^*)^2}
   + \frac{\delta m(r)}{2(m_0^*)^2}\hat{p}^2
   + 2\hat{p}\frac{\delta m(r)}{2(m_0^*)^2}\hat{p}\right) .
\end{equation}
The necessity of application of Weyl ordering in mathematically similar problems 
has been considered in 
\cite{PhysRevA.23.1591,PhysRevA.34.3565,PhysRevA.38.3101}. As it was pointed 
in \cite{PhysRevA.89.032111}, in the FW transformed Dirac-Pauli equations the 
``operators orderings is a matter of further investigations and particularly, 
considerable insight might be obtained by exploiting the analogy to the model of 
semiconductor heterostructures''.

One additional remark is necessary. As follows from above there are no any 
ordering problems as long as our consideration is carried out within momentum 
space. All problems arise in the configurational space. It is one more 
advantage of this representation showed up in the cause of proposed approach to 
the reduction of multicomponent $\vec{k} \cdot \vec{p}$ Hamiltonians. This 
contradicts the statement in \cite{Reiher2006} that reference to momentum space 
is ``myth''. Using just this representation, we can bypass not only the ordering 
problem but can ``consider the boundary conditions at interfaces in a natural 
way, and avoid automatically any spurious solutions'' in quite general  
Hamiltonian as it has been argued in \cite{Winkler1993}.

In the special case $E_g = 0$ (graphene) external potential 
has rather simple form for electrons in no-pair approximation
\begin{equation}
V_e = \frac{1}{2}\int\int\left( 1 + e^{-i\varDelta(\vec{k},\vec{q})} 
\right)
 V(\vec{k}-\vec{q})\hat{a}^+(\vec{k})\hat{a}(\vec{q})
 \small{\frac{d\vec{k} d\vec{q}}{(2\pi)^4}}.
\end{equation}
It is seen that the terms modifying scattering potential in the single particle 
channel are of the order of bare potential and do not depend on semiconductor 
parameters. Thus, generally their effect cannot be considered as perturbation 
even in single particle channel. One of the consequences is, that in the single 
particle scattering channel (as for electrons, so for holes) the back-scattering 
processes ($\varDelta(\vec{k},\vec{q}) = \pi$) are strictly suppressed. The 
modified Pauli potential $V_e \equiv 0$ for any chosen external scalar 
potential in this case. Here we have also correspondence with a well-known 
result in graphene physics, which is the origin of arising of topologically 
protected surface states \cite{Ando1998,PhysRevLett.102.096801}. The most vivid 
example of this effect according to formulated equations is electron kinetics in 
1D ($k_y = 0$). As in this case  $\varDelta(k_x, q_x)$ takes the $0$, $\pi$ 
values only, the interaction with impurity (external) potential is
\begin{eqnarray}
V_e&=&\int\int\left(\varTheta(k_x)\varTheta(q_x)
 + \varTheta(-k_x)\varTheta(-q_x) \right) \nonumber \\
&\times& V(k_x-q_x)\hat{a}^+(k_x)\hat{a}(q_x)\frac{dk_x dq_x}{(2\pi)^2},
\label{eq_Ve_1D}
\end{eqnarray}
where $\varTheta(x)$ is a Heaviside step function. It is seen that in the 
electron single particle channel in 1D problem we have independent ``right''
moving solutions spanned by $\varTheta(k_x)\hat{a}(k_x)$ the and ``left'' 
moving solutions formed by $\varTheta(-k_x)\hat{a}(k_x)$. Note, that within 
considered approach this result holds as long as we do not take into account 
pair contribution. If particle number is not conserved, which 
means the admixture of the states ``particle + pairs'', this restriction can be 
removed, leading to Andreev-like multi-particle type of reflection 
\cite{Andreev1964}.

The effect of mass domain wall is also best demonstrated for $E_{g0}=0$ in 1D. 
(For example, this situation can be considered as the boundary of 
heterostructure CdTe/HgTe). In this case $\varDelta(\vec{k},\vec{q}) = 0, \pi$ 
and the corresponding interaction is given by
\begin{eqnarray}
V_e&=&\int\int\left(\varTheta(k_x)\varTheta(-q_x)
 + \varTheta(-k_x)\varTheta(q_x) \right) \nonumber \\
&\times& \delta E_g(k_x-q_x)\hat{a}^+(k_x)\hat{a}(q_x)\frac{dk_x 
dq_x}{(2\pi)^2}.
\end{eqnarray}
The same expression is valid for holes. As compare with potential scattering 
\eqref{eq_Ve_1D}, lattice distortion favors back scattering and thus can lead 
to localization at such boundary as for electrons, so for holes.

Returning to the effect of potential modification in graphene, we must add that 
the noted possible spatial separation of electrons and holes by applied external 
perturbation is in accord with the results of \cite{Zhang2009}. In this paper 
the authors reported a new technique of Dirac point mapping of charge 
inhomogeneities in graphene. Using it, they have revealed that charge puddles 
are caused by charge-donating impurities below the graphene. At the same time 
they have ruled out the hypothesis that topographic corrugations in graphene 
were a primary cause of the charge separation. This result is also in accord 
with our conclusion that the effect of lattice distortions upon electrons and 
holes is alike.

To investigate further properties of modified potential in no-pair equations let 
us write it down for the linear potential $V(\vec{r}) = F \cdot x$ chosen 
in momentum representation as
\begin{eqnarray}
\label{eq_V_lin}
  V(\vec{k} - \vec{q})
   &=& F \cdot \frac{i}{2} \left\lbrace
                    \delta_{k_x}'(\vec{k} - \vec{q})
                  - \delta_{q_x}'(\vec{q} - \vec{k})
                  \right\rbrace \\
   &=& F \cdot \delta_{k_x, q_x}'(\vec{k}, \vec{q}) \ .
\end{eqnarray}
Substituting \eqref{eq_V_lin} into \eqref{Dirac_Ve} we obtain two terms: one 
``trivial'' term $iF\frac{\partial}{\partial k_x} \varphi(k)$ describing 
interaction with ``bare'' potential and terms depending only on momentums which 
can be attributed to kinetic energy. The modified effective kinetic term in 
this case is
\begin{eqnarray}
  \nonumber
  E_{kin,eff} &=& \varepsilon(\vec{k})
          \pm F \frac{\gamma^2 k_y}{\varepsilon(k)(E_g + 2\varepsilon(k))} \\
  \label{eq_kin_eff}
 &=& \varepsilon(\vec{k}) \pm \frac{1}{2}F\frac{\partial}{\partial k_y}
 \ln (1 + 2 \varepsilon(\vec{k})/E_g).
\end{eqnarray}
The different signs refer to two group of states of the problem. It follows 
from \eqref{eq_kin_eff} that in EFA  this expression acquires the customary 
Rashba form \cite{Rashba1984}
\begin{equation}
  E_{kin,eff} \approx \frac{\gamma^2 k^2}{E_g} \pm 
       \frac{F \gamma^2}{E_g^2}k_y,
\end{equation}
where $F \gamma^2/E_g^2$ plays the role of effective Rashba coupling constant.

If the electric field is applied along Y axis the Rashba term will be
\begin{equation}
\mp \frac{1}{2}F\frac{\partial}{\partial k_x}
 \ln (1 + 2 \varepsilon(\vec{k})/E_g).
\end{equation}
Thus in considered approach Rashba effect can be considered as due to some 
Berry-like field acting in momentum space with connection
\begin{equation}
 \vec{A}(\vec{k}) = (\gamma^2 k_x/f(k), -\gamma^2 k_y/f(k)),
\end{equation}
where $f(k) = \varepsilon(k)(E_g + 2\varepsilon(k))$. The corresponding 
Berry-like curvature has the most simple form in EFA
\begin{equation}
 \Omega_z \approx \frac{\gamma^2}{E_g^2} = \lambda_C^2.
\end{equation}
The statement that arising of Rashba term is due to Berry-like field will be 
confirmed additionally below while considering position operator within SQM. 

\section{Position operator in 2D Dirac-like semiconductor}
\label{sec:3}
Let us consider obtained results from another point of view. We will be 
interested in the form and properties of position operator $\hat{\vec{r}}$ in 
SQM. Using the definition
\begin{eqnarray}
  \nonumber
\hat{\vec{r}} 
&=& \sum_{s=1,2}\int\int \Psi_s^+(\vec{k})\delta_{k_x, q_x}'(\vec{k}, 
\vec{q})\Psi_s(\vec{q})
  \frac{d\vec{k}}{(2\pi)^2}\frac{d\vec{q}}{(2\pi)^2} \\
&=& \sum_{s=1,2} \int \hat{\vec{r}}_s(\vec{k}) \frac{d\vec{k}}{(2\pi)^2},
\end{eqnarray}
we obtain
\begin{equation}
\label{eq_R_general}
 \hat{\vec{r}}(\vec{k}) = \sum_{s=1,2}\left( 
  \hat{\vec{R}}_s(\vec{k}) + \hat{\vec{A}}_s(\vec{k}) + \hat{\vec{B}}_s(\vec{k})
  \right).
\end{equation}
Here
\begin{equation}
 \hat{\vec{R}}_s(\vec{k}) = 
  \hat{\vec{r}}_{Sch}(\vec{k}) \left(
   \hat{a}_s^+(\vec{k})\hat{a}_s(\vec{k}) - 
\hat{b}_s^+(\vec{-k})\hat{b}_s(\vec{-k}) 
\right).
\end{equation}
In this expression $\hat{\vec{r}}_{Sch}(\vec{k})$ is customary Schrodinger 
position operator $i d/d\vec{k}$ in momentum space when acting 
in single particle channel on corresponding wave function. Here we remember that 
in general, the Dirac problem involves two group of states labeled by index 
$s=1,2$. The account of second pair of states will be required below while 
analyzing the origin of Darwin-like terms. As regards immediately ensuing 
discussion, this account is of no consequence and we shall omit index  for the 
time being. 
\begin{gather}
 \hat{\vec{A}}(\vec{k}) = \vec{A}(\vec{k}) \left(
   \hat{a}^+(\vec{k})\hat{a}(\vec{k}) + \hat{b}^+(\vec{-k})\hat{b}(\vec{-k}) 
\right), \\
 \vec{A}(\vec{k}) = - \frac{1}{2}[\vec{n}_z \times \vec{\nabla}_{\vec{k}}]
   \ln (E_g + 2 \varepsilon(\vec{k})). \label{eq_A_k}
\end{gather}
The term $\hat{\vec{B}}(\vec{k})$ is responsible for pair participation 
processes.
\begin{gather}
\label{eq_Pos_oper_pairs}
\hat{\vec{B}}(\vec{k}) = 
 \vec{B}(\vec{k})\hat{a}^+(\vec{k})\hat{b}^+(\vec{-k})
 + \vec{B}^*(\vec{k})\hat{b}(\vec{-k})\hat{a}(\vec{k}) \\
 \vec{B}(\vec{k}) = \left( i \frac{\partial\varTheta(\vec{k})}{\partial\vec{k}}
 + \frac{\sin 
2\varTheta(\vec{k})}{2}\frac{\partial\varphi(\vec{k})}{\partial\vec{k}}
 \right) e^{-i\varphi(\vec{k})}
\end{gather}
In SQM approach the time dependence of position operator in Heisenberg 
representation in ``free'' problem is determined solely by the time dependence 
of creation/annihilation operators. It means that any time-dependent 
oscillating effects with frequency proportional to gap width, affecting single 
particle wave packets evolution, are due to virtual pair participation (see 
\eqref{eq_Pos_oper_pairs}). This result is in accord with the statement: ``In 
terms of condensed matter physics, the Zitterbewegung is nothing but a special 
kind of inter-band transitions with creation of virtual electron-hole pairs'' 
\cite{Katsnelson2006}. Thus in no-pair approximation, which is true at least 
far from scattering centers, the time-dependent self-smearing of electron is 
absent \cite{PhysRevA.72.064103}.

From \eqref{eq_A_k} it follows that $\hat{\vec{A}}(\vec{k})$ terms can be 
considered as due to some 
$rot$ field which reminds Berry connection behavior. We will add always the 
word ``like'' while referring to this field as in our case it does not 
possesses the topological properties of ``real'' Berry phase considered e.g. in 
the Bloch periodic problem \cite{PhysRevLett.62.2747} due to unboundedness of 
the spectrum in considered $\vec{k} \cdot \vec{p}$ problem.

The projections of position operator for the first group of states into single 
electron channel in no-pair approximation dependent only on electron number 
operators are
\begin{equation}
\label{eq_Pos_oper_x}
  \hat{x}(\vec{k}) = 
  \left[ \hat{x}_{sch}(\vec{k}) + 
         \frac{1}{2}\frac{\partial}{\partial k_y}
           \ln (E_g + 2 \varepsilon(\vec{k}))\right]
   \hat{a}^+(\vec{k}) \hat{a}(\vec{k}),
\end{equation}
where $\hat{x}_{sch}(\vec{k}) = i \partial / \partial k_x$. Similarly we 
can write down the following expression for $\hat{y}(\vec{k})$
\begin{equation}
  \hat{y}(\vec{k}) = 
  \left[ \hat{y}_{sch}(\vec{k}) - 
         \frac{1}{2}\frac{\partial}{\partial k_x}
           \ln (E_g + 2 \varepsilon(\vec{k}))\right]
   \hat{a}^+(\vec{k}) \hat{a}(\vec{k}),
\end{equation}
The coordinates in momentum representation satisfy the deformed Heisenberg 
algebra in single particle channel \cite{PhysRevD.69.127701}
\begin{equation}
 \left[ \hat{x}(\vec{k}), \hat{y}(\vec{k}) \right] = 
   -i \frac{1}{2} \bigtriangleup \ln (E_g + 2 \varepsilon(\vec{k}))
       \hat{a}^+(\vec{k}) \hat{a}(\vec{k}).
\end{equation}
This result of cause is not new. It was first postulated by Bacry 
\cite{Bacry1988}. In \cite{Berard2006190} this anomalous 
contribution to position operator in momentum space has been obtained and 
discussed within FW projection onto positive energy states.

For $E_g = 0$ position operators are simplified to
\begin{gather}
\hat{x}(\vec{k}) = 
  \left[ \hat{x}_{sch}(\vec{k}) + \frac{k_y}{2k^2} \right]
     \hat{a}^+(\vec{k}) \hat{a}(\vec{k}), \\
\hat{y}(\vec{k}) = 
  \left[ \hat{y}_{sch}(\vec{k}) - \frac{k_x}{2k^2} \right]
     \hat{a}^+(\vec{k}) \hat{a}(\vec{k}).
\end{gather}
Their commutator is of pure ``monopole'' type
\begin{equation}
 \left[ \hat{x}(\vec{k}), \hat{y}(\vec{k}) \right] = 
   -i \frac{1}{2} \delta(\vec{k})\hat{a}^+(\vec{k}) \hat{a}(\vec{k}).
\end{equation}

The presented form of position operators contradict the statement that 
Zitterbewegung ``has a close relation to Berry connection'' 
\cite{PhysRevB.81.121417}. Within proposed consideration in no-pair 
approximation the Berry field carrying term in defined position operators does 
not include the contribution of the pair creation processes and thus does not 
depend on time. This Berry field cannot be the origin of time-oscillation 
phenomenon, which is the hallmark of Zitterbewegung. 

Now the idea is to replace Schrodinger operator $\hat{\vec{r}}_{Sch}(\vec{k}) = 
i d / d\vec{k}$ in Schrodinger expression for interaction with external 
potential by these Berry-like phase carrying operators. To this end, we use the 
following chain of equalities
\begin{gather}
\nonumber
\int\frac{d\vec{q}}{(2\pi)^2}V(\vec{k} - \vec{q})\varphi(\vec{q})
= \int\frac{d\vec{Q}}{(2\pi)^2}V(-\vec{Q})\varphi(\vec{k}+\vec{Q}) \\
\nonumber
= \left[ \int \frac{d\vec{Q}}{(2\pi)^2}V(-\vec{Q})
 e^{i\hat{\vec{k}}\vec{Q}}\right]\varphi(\vec{k}) \\
= \left[ \int \frac{d\vec{Q}}{(2\pi)^2}V(-\vec{Q})
 e^{-i\hat{\vec{r}}_{Sch}(\vec{k})\vec{Q}}\right]\varphi(\vec{k}) .
 \label{eq_R_subst}
\end{gather}
Here $\hat{\vec{k}} = -id / d\vec{k}$ and substitution is carried out in the 
last expression.

In order to simplify estimations and present the main result more clear we will 
again restrict ourselves to EFA. Within this approximation
\begin{equation}
 \vec{A}(\vec{k}) \approx \frac{2\gamma^2}{E_g^2}(k_y, -k_x), \:
 \vec{B}(\vec{k}) \approx \frac{\gamma}{E_g}(i, 1).
\end{equation}
In this case the position operators commutator is
\begin{equation}
 \left[ \hat{x}(\vec{k}), \hat{y}(\vec{k}) \right] = 
   -i \frac{\gamma^2}{E_g^2}\hat{a}^+(\vec{k}) \hat{a}(\vec{k}).
\end{equation}
It is in one to one correspondence with the commutator of velocity components 
in the system under constant magnetic field. Now let us replace 
$\hat{r}_{sch}(\vec{k})$ in \eqref{eq_R_subst} by operators $\hat{r}(\vec{k})$
formed by $\vec{A}(\vec{k})$ and $\vec{B}(\vec{k})$. Using frequently employed 
in physical problems equation
\begin{equation}
 e^A e^B = e^{A+B+C/2},
\end{equation}
which is true if the commutator $C$ of $A$ and $B$ commutes with both $A$ and 
$B$. The $\exp(-i\hat{\vec{r}}(\vec{k})\vec{Q})$ can be written as
\begin{eqnarray}
\nonumber
 e^{-i\hat{\vec{r}}(\vec{k})\vec{Q}} &=&
 e^{-i\frac{2\gamma^2}{E_g}k_x Q_y} e^{i\frac{2\gamma^2}{E_g}k_y Q_x}
   e^{i\hat{\vec{r}}_{Sch}(\vec{k})\vec{Q}} \\
 &\approx& \left[ 1 -  \frac{2\gamma^2}{E_g}(k_x Q_y - k_y Q_x))\right]
  e^{i\hat{\vec{r}}_{Sch}(\vec{k})\vec{Q}}.
\end{eqnarray}
It is easily verified that inserting this expression into \eqref{eq_R_subst} 
will modify external potential adding spin-orbit-type term (compare with 
\eqref{eq110}) but only this term.

The $\vec{k} \cdot \vec{p}$ Berry phase carrying part of position operator and 
only it has a hand in generation of SOI-type terms in no-pair equations. Now 
the natural question arises: Where Darwin contribution vanishes? It is absent in 
presented derivation despite the fact that in the low energy regime (EFA) it is 
of the same order in $\gamma k/E_g$ as spin-orbit one. In order to restore 
the true account for all terms of the second order we are to return to the 
general expression for $\hat{\vec{r}}(\vec{k})$ \eqref{eq_R_general}.

In EMA the first two terms in \eqref{eq_R_general} commute with pair-creating 
terms \eqref{eq_Pos_oper_pairs}. The decomposion of
$exp(\sum_{s=1,2} \hat{\vec{r}}_s(\vec{k})\vec{Q})$   up to the second order 
in $\gamma k/E_g$ is based on the perturbation formula
\begin{gather}
e^{(a+\lambda b)t} = \sum_{n=0}\lambda^n u_n \\
u_n = \left[\int_0^t dt_1 \int_0^{t_2}dt_2
\ldots \int_0^{t_{n-1}}dt_n b(t1)\ldots b(t_n) \right]e^{at} \\
b(t) = e^{at}be^{-at}.
\end{gather}
When acting in single particle channel in the first group of states, only the 
following contribution of the pair creation terms from the second group of 
states remains
\begin{equation}
 \frac{\gamma^2}{E_g^2}\int_0^Q \vec{B}_2^*(\vec{k})d\vec{Q_1}
 \int_0^{Q_1} \vec{B}_2(\vec{k})d\vec{Q_2} = \frac{\gamma^2}{2E_g}Q^2.
\end{equation}
Here $\vec{B}_1(\vec{k}) = \vec{B}_2^*(\vec{k})$. It is easy to check that 
addition of this term into the expression for modified potential leads to Darwin 
expression.

There is a striking difference in pure Dirac theory and presented consideration 
of semiconductor problem. It concerns the definition and ``reality'' of vacuum 
state and interpretation of downward/upward transitions in both cases 
\cite{PhysRevA.70.054101}. In intrinsic semiconductor filled 
valence bands constitutes vacuum in SQM. Electrons occupy all states with 
negative energies and Pauli principle forbids any downward transitions. The 
upward transitions are interpreted as the occurrence of quasiparticle with 
positive mass and positive charge i.e. hole. This ``hole theory'' in 
relativistic Dirac theory has been long ago reexamined. The appropriate 
interpretation considers the negative Dirac continuum as charge-conjugated 
states of positrons with positive energy \cite{Feynman1949,Schweber1962}. 
Nevertheless, despite of these differences, some quasi-relativistic effects in 
semiconductors are of the same nature. In discussed above Darwin effect we see 
that while external potential contains pair creation terms involving both types 
of states, only those terms that are due to states which are orthogonal to 
considered in single particle channel participate in the effect. This 
is analogous to the effect considered within quantum field-theoretical 
simulations accompanied by analytical estimates predicting the suppression of 
pair production at the barrier by incoming electron due to Pauli
principle\cite{PhysRevA.72.064103}.

The presented derivation poses the question about commonly used interpretation 
of spin-orbit-type and Darwin terms. In atoms, e.g., it is interpreted as 
electromagnetic interaction of self-induced magnetic field due to electron 
orbital motion with self-magnetic moment due to spin \cite{Rel_Quant_Theory}. In 
our approach this modification can be attributed to Pauli prohibition for 
electrons to scatter in occupied valence states and is by construction analogous 
to some pseudo potential used in atomic and solid state physics. As regards 
Darwin term, from the point of quantum field theory it emerges because any 
amplitude is accompanied by the amplitude for vacuum to remain vacuum - i.e. by 
the bubble diagrams representing creation and subsequent annihilation of 
electron-hole pairs from the vacuum \cite{Holstein1998}. In considered low 
energy regime we restrict our consideration to single bubble approximation. The 
fact that Darwin term has no classical correspondence was underlined 
in \cite{PhysRevA.89.032111}.

It has been shown in \cite{Berard2006190} that position algebra becomes 
non-commutative after FW transformation of Dirac equation due to Berry phase 
contribution. In our case, the Berry connection contribution coincides with the 
expression obtained in \cite{Berard2006190} if no-pair assumption is valid
\begin{equation}
 \vec{A}_{e,h} = - \frac{1}{2}[\vec{n}_z \times \vec{\nabla}]
   \ln (E_g + 2 \varepsilon(\vec{k})).
\end{equation}
The corresponding Berry curvature is
\begin{equation}
 \Omega_z = \frac{1}{2}\bigtriangleup \ln (E_g + 2 \varepsilon(\vec{k})).
\end{equation}

\begin{figure}
\centering
\includegraphics[width=.45\textwidth]{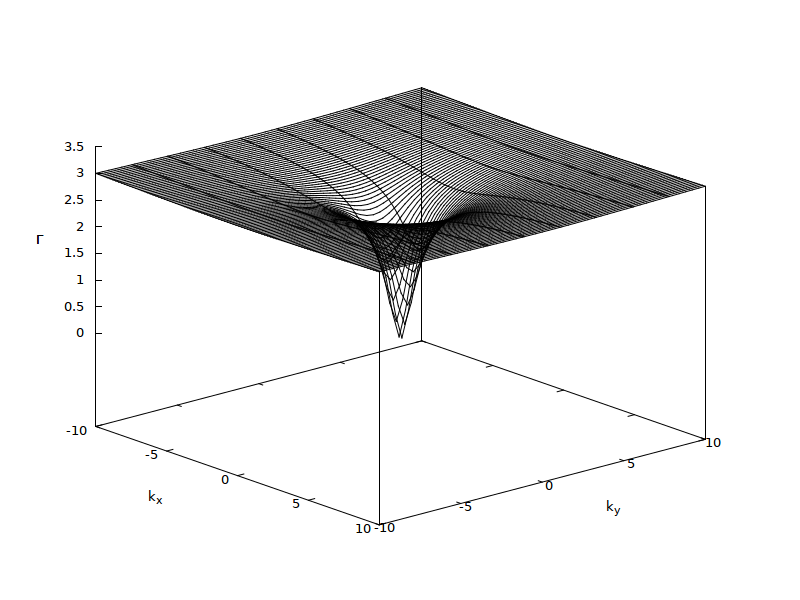}
\includegraphics[width=.45\textwidth]{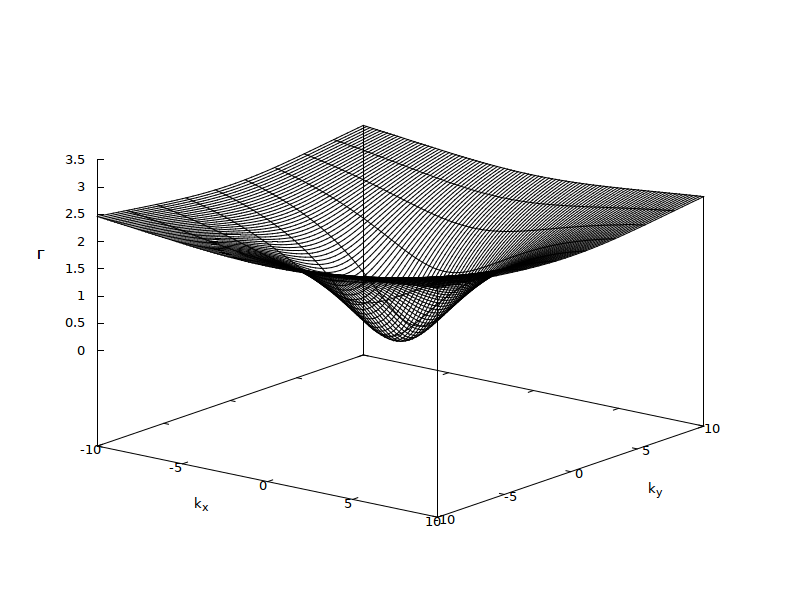}
\caption{Dependence of Berry phase for constant radius of closed path of 
integration for $E_g=10\text{ mEv}$ (top) and $E_g=50\text{ mEv}$ (bottom), 
$\gamma = 8\cdot 10^{-8} \text{mEv}\cdot\text{cm}$}
\label{Fig_Berry_Dirac_3d}
\end{figure}

\begin{figure}
\centering
\includegraphics[width=.45\textwidth]{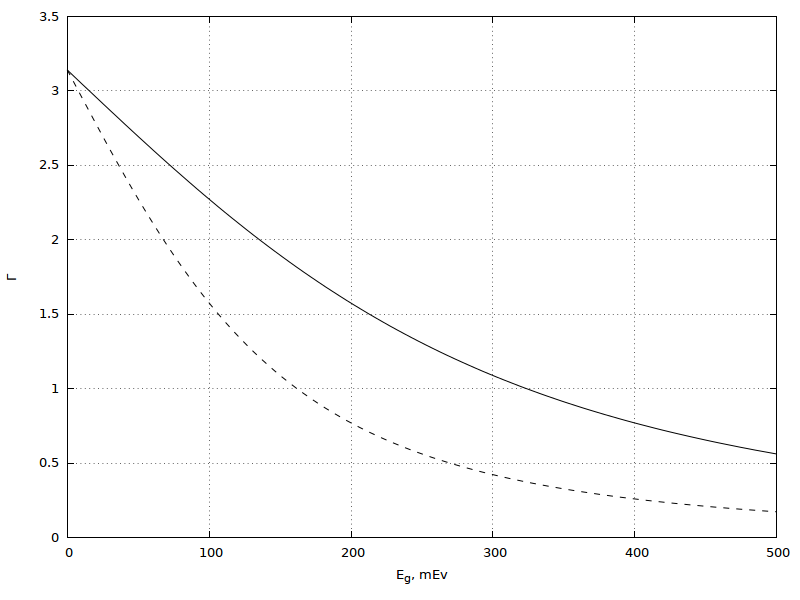}
\caption{Dependence of Berry on $E_g$ for constant radius of contour: 
$k=10^8 \text{cm}^{-1}$ (dashed line) and $k=3 \cdot 10^8 \text{cm}^{-1}$ 
(solid line), $\gamma = 8\cdot 10^{-8} \text{mEv}\cdot\text{cm}$}
\label{Fig_Berry_Dirac_Eg}
\end{figure}

There is a difference in the role of considered $\vec{k} \cdot \vec{p}$ Berry 
phase $\varGamma_e$ and Zak Berry phase defined for the dynamics of electrons 
in periodic solids \cite{PhysRevLett.62.2747}. The latter phase is geometrical 
phase that characterizes the topological properties of given Bloch band. The Zak 
phase, gained during adiabatic motion across Brillouin zone, is an invariant.
In considered  problems the spectrum is unbounded. It means that 
in general this invariant property is lost. We can analyze instead the 
dependence of general expression for Berry curvature on closed path radius $k$ 
and $E_g$
\begin{eqnarray}
  \nonumber
  \varGamma_e &=& \oint \vec{A}_{e,h}(\vec{k}) d\vec{k}
   = \int \vec{B}(\vec{k}) d\vec{S} \\
   &=& 2\pi \frac{\gamma^2 k^2}{\varepsilon(k)(E_g + 2\varepsilon(k))}. 
\end{eqnarray}
It is seen that our $\vec{k} \cdot \vec{p}$ Berry phase exists as topological 
characteristic only for $E_g=0$, when it does not depend on the chosen 
closed path radius and is equal to $\pi$. For $E_g \neq 0$ Berry phase for given 
contour decreases from $\pi$ to zero with $E_g$ increase. The dependence of 
Berry phase on $k$ (radius of closed path of integration) for two different 
values of $E_g$ is presented in Fig.\ref{Fig_Berry_Dirac_3d}. In addition, the 
dependence of Berry phase on $E_g$ for constant radius of contour ($k = const$) 
is presented in Fig.\ref{Fig_Berry_Dirac_Eg}.
The similar behavior of Berry phase was obtained in \cite{Urru2015} in the 
tight-binding model of graphene when different on-site terms are added to the 
two sublattices making them nonequivalent. In our case the considered Berry 
phase is an ``intrinsic'' property of free $\vec{k} \cdot \vec{p}$ problem. Due 
to the symmetry of the bands in Dirac problem all results for electrons holds 
for holes. 

All above considerations were carried out in momentum space. Only in this 
representation the unambiguous separation of states carrying opposite charges is 
possible. The defined ``no-pair'' position operator \eqref{eq_Pos_oper_x} is 
local in momentum representation. It means that transformation of 
$\hat{x}(\vec{k})$ to coordinate space makes it non-local operator $\hat{X}$.
Its action upon electron wave function $\psi(x, k_y)$ is defined as
\begin{eqnarray}
  \nonumber
\hat{X}\psi(x, k_y) &=& \int [\hat{x}(\vec{k})\psi(k_x, k_y)]e^{ik_xx}dk_x \\
  \nonumber
&=& \int\int \psi(x', k_y)[\hat{x}(\vec{k})e^{-ik_xx'}]e^{ik_xx}dk_x dx' \\
\nonumber
&=& x\psi(x, k_y) + \int K(x-x', k_y)\psi(x', k_y)dx'.
\end{eqnarray}
The kernel $K(x-x', k_y)$  determining non-local behavior of position operator 
is
\begin{equation}
 K(\Delta x, k_y) =
 \frac{1}{2} \int \frac{\partial}{\partial k_y} \ln (E_g + 2 
\varepsilon(\vec{k}))
  e^{ik_x\Delta x}dk_x.
\end{equation}
In the limit $\gamma k/E_g \ll 1$ the asymptotical behavior of the kernel
$K(\Delta x, k_y)$ is
\begin{equation}
  K(\Delta x, q_y) \sim \frac{\gamma k_y}{E_g}
   \left( \sqrt{2}e^{-\frac{E_g}{\sqrt{2}\gamma}|\Delta x|}
     - e^{-\frac{E_g}{\gamma}|\Delta x|} \right) ,
\end{equation}
The smearing is determined by ``Compton'' wave length $\gamma/E_g$. In the
opposite limit $E_g \rightarrow 0$
\begin{equation}
  K(\Delta x, k_y) \sim e^{-|k_y||\Delta x|} ,
\end{equation}
i.e. the smearing is determined by de Broglie wave length $1/k_y$. 
The non-locality of position operator was predicted for the relativistic 
particles in famous paper by T.D. Newton and E.P.Wigner 
\cite{RevModPhys.21.400}. The authors has shown that for elementary system with 
relativistic energy dispersion relation $E(k)=\sqrt{E_g^2+\gamma^2
k^2}$ the position operator becomes integral operator in position space instead 
of common multiplicative c-number operator. As it follows from our consideration 
this effect appears as long as we account for filled valence band in our 
multiband problem. The same effect was discussed in 
\cite{PhysRevB.72.085217,Zawadzki2011,Zawadzki2012}, where it
was stated that the electrons in narrow gap semiconductors are to be
considered as the extended objects of size $\lambda\sim \hbar(m^\star
P)$ ($m^\star$ is the effective mass). It was pointed out that in
narrow gap semiconductors $\lambda$ can be as large as $70$~\AA.

\section{Summary}
The derivation of Pauli-like single-particle no-pair equations in intrinsic 2D 
Dirac-like semiconductors was carried out within SQM. The truncation of $\vec{k} 
\cdot \vec{p}$ Hamiltonian to single particle channels leads to essential 
modification of external potential and lattice distortion perturbation, 
especially in degenerate gapless situation. We confirm the validity of proposed 
approach by comparison of obtained results with the well-known and 
experimentally established ones. The similar Pauli equations for electrons were 
obtained and discussed using Casimir-type positive energy projection operators 
\cite{PhysRevA.30.703}. We demonstrated the advantage of proposed approach over 
FW truncation and its variants. For example, as it has been proven in 
\cite{Neznamov2009} the Foldy-Wouthousen transformation must be accompanied by 
nullification of either upper or lower components of bispinor wave function. In 
our approach this result immediately follows in no-pair approximation. The SQM 
approach presents simple and clear physical picture formulating single particle 
equations for both types of carriers. The difference in a role played by single 
particle processes and participation of virtual pairs is revealed. SQM approach 
allows to take into account the effect of filled valence bands and to avoid in 
the cause of derivation unphysical superposition of positive and negative energy 
states. The account for filling modifies essentially external potential and 
lattice distortion perturbation entering the sought equations making them in 
general non-local operators. This modification does not depend on semiconductor 
parameters in the degenerate case and is determined by perturbation strength 
only. In our approach, the effective single-particle potential is constructed in 
such a way as to exclude scattering of electrons into occupied valence states. 
Remaining faithful to Pauli principle, we obtain on this way some kind of pseudo 
potential, which actually does not differ in essence from the one used in solid 
state physics. The additional insight into origin of potential modification is 
provided by analyzing properties of position operator in SQM. The results of 
\cite{Zhang2009} suggest that unobservable relativistic effects can be 
successfully reproduced in the systems with more ``user friendly'' parameters 
\cite{PhysRevA.86.032103}. We believe that the proposed approach can serve as 
theoretical grounds for comparison of quasi-relativistic effects in 
semiconductors with numerical solutions to relativistic QFT with space-time 
resolution \cite{PhysRevA.72.064103}.

\acknowledgments
We are grateful to A.V.~Germanenko and G.M.~Minkov for stimulating
discussions. The work has been supported in part by the RFBR (Grant No.
13-02-00322) and the Ministry of education and science of Russia
(Project No. 3.571.2014/K).


\begin{thebibliography}{59}%
\makeatletter
\providecommand \@ifxundefined [1]{%
 \@ifx{#1\undefined}
}%
\providecommand \@ifnum [1]{%
 \ifnum #1\expandafter \@firstoftwo
 \else \expandafter \@secondoftwo
 \fi
}%
\providecommand \@ifx [1]{%
 \ifx #1\expandafter \@firstoftwo
 \else \expandafter \@secondoftwo
 \fi
}%
\providecommand \natexlab [1]{#1}%
\providecommand \enquote  [1]{``#1''}%
\providecommand \bibnamefont  [1]{#1}%
\providecommand \bibfnamefont [1]{#1}%
\providecommand \citenamefont [1]{#1}%
\providecommand \href@noop [0]{\@secondoftwo}%
\providecommand \href [0]{\begingroup \@sanitize@url \@href}%
\providecommand \@href[1]{\@@startlink{#1}\@@href}%
\providecommand \@@href[1]{\endgroup#1\@@endlink}%
\providecommand \@sanitize@url [0]{\catcode `\\12\catcode `\$12\catcode
  `\&12\catcode `\#12\catcode `\^12\catcode `\_12\catcode `\%12\relax}%
\providecommand \@@startlink[1]{}%
\providecommand \@@endlink[0]{}%
\providecommand \url  [0]{\begingroup\@sanitize@url \@url }%
\providecommand \@url [1]{\endgroup\@href {#1}{\urlprefix }}%
\providecommand \urlprefix  [0]{URL }%
\providecommand \Eprint [0]{\href }%
\providecommand \doibase [0]{http://dx.doi.org/}%
\providecommand \selectlanguage [0]{\@gobble}%
\providecommand \bibinfo  [0]{\@secondoftwo}%
\providecommand \bibfield  [0]{\@secondoftwo}%
\providecommand \translation [1]{[#1]}%
\providecommand \BibitemOpen [0]{}%
\providecommand \bibitemStop [0]{}%
\providecommand \bibitemNoStop [0]{.\EOS\space}%
\providecommand \EOS [0]{\spacefactor3000\relax}%
\providecommand \BibitemShut  [1]{\csname bibitem#1\endcsname}%
\let\auto@bib@innerbib\@empty
\bibitem [{\citenamefont {Winkler}(2003)}]{Winkler_SO_Coupl_eff}%
  \BibitemOpen
  \bibfield  {author} {\bibinfo {author} {\bibfnamefont {R.}~\bibnamefont
  {Winkler}},\ }\href@noop {} {\emph {\bibinfo {title} {Spin-Orbit Coupling
  Effects in Two-Dimensional Electron and Hole Systems}}}\ (\bibinfo
  {publisher} {Springer},\ \bibinfo {address} {Berlin},\ \bibinfo {year}
  {2003})\BibitemShut {NoStop}%
\bibitem [{\citenamefont {Katsnelson}\ and\ \citenamefont
  {Novoselov}(2007)}]{Katsnelson20073}%
  \BibitemOpen
  \bibfield  {author} {\bibinfo {author} {\bibfnamefont {M.}~\bibnamefont
  {Katsnelson}}\ and\ \bibinfo {author} {\bibfnamefont {K.}~\bibnamefont
  {Novoselov}},\ }\href {\doibase http://dx.doi.org/10.1016/j.ssc.2007.02.043}
  {\bibfield  {journal} {\bibinfo  {journal} {Solid State Communications}\
  }\textbf {\bibinfo {volume} {143}},\ \bibinfo {pages} {3 } (\bibinfo {year}
  {2007})}\BibitemShut {NoStop}%
\bibitem [{\citenamefont {Rusin}\ and\ \citenamefont
  {Zawadzki}(2011)}]{Zawadzki2011}%
  \BibitemOpen
  \bibfield  {author} {\bibinfo {author} {\bibfnamefont {T.~M.}\ \bibnamefont
  {Rusin}}\ and\ \bibinfo {author} {\bibfnamefont {W.}~\bibnamefont
  {Zawadzki}},\ }\href {\doibase 10.1103/PhysRevA.84.062124} {\bibfield
  {journal} {\bibinfo  {journal} {Phys. Rev. A}\ }\textbf {\bibinfo {volume}
  {84}},\ \bibinfo {pages} {062124} (\bibinfo {year} {2011})}\BibitemShut
  {NoStop}%
\bibitem [{\citenamefont {Zawadzki}\ and\ \citenamefont
  {Rusin}(2011)}]{Zawadzki2011_ZB}%
  \BibitemOpen
  \bibfield  {author} {\bibinfo {author} {\bibfnamefont {W.}~\bibnamefont
  {Zawadzki}}\ and\ \bibinfo {author} {\bibfnamefont {T.~M.}\ \bibnamefont
  {Rusin}},\ }\href@noop {} {\bibfield  {journal} {\bibinfo  {journal} {J.
  Phys.: Condens. Matter}\ }\textbf {\bibinfo {volume} {23}},\ \bibinfo {pages}
  {143201} (\bibinfo {year} {2011})}\BibitemShut {NoStop}%
\bibitem [{\citenamefont {Katsnelson}\ and\ \citenamefont
  {Geim}(2006)}]{Kats06}%
  \BibitemOpen
  \bibfield  {author} {\bibinfo {author} {\bibfnamefont {K.}~\bibnamefont
  {Katsnelson}, \bibfnamefont {M~Novoselov}}\ and\ \bibinfo {author}
  {\bibfnamefont {A.}~\bibnamefont {Geim}},\ }\href@noop {} {\bibfield
  {journal} {\bibinfo  {journal} {Nature Physics}\ }\textbf {\bibinfo {volume}
  {2}},\ \bibinfo {pages} {620} (\bibinfo {year} {2006})}\BibitemShut {NoStop}%
\bibitem [{\citenamefont {Keldysh}(1963)}]{Keldysh1963}%
  \BibitemOpen
  \bibfield  {author} {\bibinfo {author} {\bibfnamefont {L.~V.}\ \bibnamefont
  {Keldysh}},\ }\href@noop {} {\bibfield  {journal} {\bibinfo  {journal}
  {JETP}\ }\textbf {\bibinfo {volume} {45}},\ \bibinfo {pages} {365} (\bibinfo
  {year} {1963})}\BibitemShut {NoStop}%
\bibitem [{\citenamefont {Foldy}\ and\ \citenamefont
  {Wouthuysen}(1950)}]{Foldy1950}%
  \BibitemOpen
  \bibfield  {author} {\bibinfo {author} {\bibfnamefont {L.~L.}\ \bibnamefont
  {Foldy}}\ and\ \bibinfo {author} {\bibfnamefont {S.~A.}\ \bibnamefont
  {Wouthuysen}},\ }\href {\doibase 10.1103/PhysRev.78.29} {\bibfield  {journal}
  {\bibinfo  {journal} {Phys. Rev.}\ }\textbf {\bibinfo {volume} {78}},\
  \bibinfo {pages} {29} (\bibinfo {year} {1950})}\BibitemShut {NoStop}%
\bibitem [{\citenamefont {Douglas}\ and\ \citenamefont
  {Kroll}(1974)}]{Douglas197489}%
  \BibitemOpen
  \bibfield  {author} {\bibinfo {author} {\bibfnamefont {M.}~\bibnamefont
  {Douglas}}\ and\ \bibinfo {author} {\bibfnamefont {N.~M.}\ \bibnamefont
  {Kroll}},\ }\href {\doibase http://dx.doi.org/10.1016/0003-4916(74)90333-9}
  {\bibfield  {journal} {\bibinfo  {journal} {Annals of Physics}\ }\textbf
  {\bibinfo {volume} {82}},\ \bibinfo {pages} {89 } (\bibinfo {year}
  {1974})}\BibitemShut {NoStop}%
\bibitem [{\citenamefont {Hess}(1986)}]{PhysRevA.33.3742}%
  \BibitemOpen
  \bibfield  {author} {\bibinfo {author} {\bibfnamefont {B.~A.}\ \bibnamefont
  {Hess}},\ }\href {\doibase 10.1103/PhysRevA.33.3742} {\bibfield  {journal}
  {\bibinfo  {journal} {Phys. Rev. A}\ }\textbf {\bibinfo {volume} {33}},\
  \bibinfo {pages} {3742} (\bibinfo {year} {1986})}\BibitemShut {NoStop}%
\bibitem [{\citenamefont {Kutzelnigg}(1990)}]{Kutzelnigg1990}%
  \BibitemOpen
  \bibfield  {author} {\bibinfo {author} {\bibfnamefont {W.}~\bibnamefont
  {Kutzelnigg}},\ }\href {\doibase 10.1007/BF01436910} {\bibfield  {journal}
  {\bibinfo  {journal} {Zeitschrift für Physik D Atoms, Molecules and
  Clusters}\ }\textbf {\bibinfo {volume} {15}},\ \bibinfo {pages} {27}
  (\bibinfo {year} {1990})}\BibitemShut {NoStop}%
\bibitem [{\citenamefont {Fearing}\ \emph {et~al.}(1994)\citenamefont
  {Fearing}, \citenamefont {Poulis},\ and\ \citenamefont
  {Scherer}}]{Fearing1994657}%
  \BibitemOpen
  \bibfield  {author} {\bibinfo {author} {\bibfnamefont {H.}~\bibnamefont
  {Fearing}}, \bibinfo {author} {\bibfnamefont {G.}~\bibnamefont {Poulis}}, \
  and\ \bibinfo {author} {\bibfnamefont {S.}~\bibnamefont {Scherer}},\ }\href
  {\doibase http://dx.doi.org/10.1016/0375-9474(94)90078-7} {\bibfield
  {journal} {\bibinfo  {journal} {Nuclear Physics A}\ }\textbf {\bibinfo
  {volume} {570}},\ \bibinfo {pages} {657 } (\bibinfo {year}
  {1994})}\BibitemShut {NoStop}%
\bibitem [{\citenamefont {Barysz}(2001)}]{Barysz2001}%
  \BibitemOpen
  \bibfield  {author} {\bibinfo {author} {\bibfnamefont {M.}~\bibnamefont
  {Barysz}},\ }\href {\doibase http://dx.doi.org/10.1063/1.1370532} {\bibfield
  {journal} {\bibinfo  {journal} {The Journal of Chemical Physics}\ }\textbf
  {\bibinfo {volume} {114}},\ \bibinfo {pages} {9315} (\bibinfo {year}
  {2001})}\BibitemShut {NoStop}%
\bibitem [{\citenamefont {Silenko}(2003)}]{Silenko2003}%
  \BibitemOpen
  \bibfield  {author} {\bibinfo {author} {\bibfnamefont {A.~J.}\ \bibnamefont
  {Silenko}},\ }\href {\doibase http://dx.doi.org/10.1063/1.1579991} {\bibfield
   {journal} {\bibinfo  {journal} {Journal of Mathematical Physics}\ }\textbf
  {\bibinfo {volume} {44}},\ \bibinfo {pages} {2952} (\bibinfo {year}
  {2003})}\BibitemShut {NoStop}%
\bibitem [{\citenamefont {Chen}\ and\ \citenamefont
  {Chiou}(2014{\natexlab{a}})}]{PhysRevA.90.012112}%
  \BibitemOpen
  \bibfield  {author} {\bibinfo {author} {\bibfnamefont {T.-W.}\ \bibnamefont
  {Chen}}\ and\ \bibinfo {author} {\bibfnamefont {D.-W.}\ \bibnamefont
  {Chiou}},\ }\href {\doibase 10.1103/PhysRevA.90.012112} {\bibfield  {journal}
  {\bibinfo  {journal} {Phys. Rev. A}\ }\textbf {\bibinfo {volume} {90}},\
  \bibinfo {pages} {012112} (\bibinfo {year} {2014}{\natexlab{a}})}\BibitemShut
  {NoStop}%
\bibitem [{\citenamefont {Liu}(2014)}]{Liu201459}%
  \BibitemOpen
  \bibfield  {author} {\bibinfo {author} {\bibfnamefont {W.}~\bibnamefont
  {Liu}},\ }\href {\doibase http://dx.doi.org/10.1016/j.physrep.2013.11.006}
  {\bibfield  {journal} {\bibinfo  {journal} {Physics Reports}\ }\textbf
  {\bibinfo {volume} {537}},\ \bibinfo {pages} {59 } (\bibinfo {year}
  {2014})},\ \bibinfo {note} {advances in Relativistic Molecular Quantum
  Mechanics}\BibitemShut {NoStop}%
\bibitem [{\citenamefont {Bir}\ and\ \citenamefont {Pikus}(1974)}]{Pikus1974}%
  \BibitemOpen
  \bibfield  {author} {\bibinfo {author} {\bibfnamefont {G.~L.}\ \bibnamefont
  {Bir}}\ and\ \bibinfo {author} {\bibfnamefont {G.~E.}\ \bibnamefont
  {Pikus}},\ }\href@noop {} {\emph {\bibinfo {title} {Symmetry and
  Strain-induced Effects in Semiconductors}}}\ (\bibinfo  {publisher} {Wiley},\
  \bibinfo {address} {New York},\ \bibinfo {year} {1974})\BibitemShut {NoStop}%
\bibitem [{\citenamefont {Pauli}(1958)}]{Pauli1958}%
  \BibitemOpen
  \bibfield  {author} {\bibinfo {author} {\bibfnamefont {W.}~\bibnamefont
  {Pauli}},\ }\href@noop {} {\emph {\bibinfo {title} {Die allgemeinen
  Prinzipien der Wellenmechanik}}}\ (\bibinfo  {publisher} {Springer},\
  \bibinfo {address} {Berlin},\ \bibinfo {year} {1958})\BibitemShut {NoStop}%
\bibitem [{\citenamefont {Berestetskii}\ \emph {et~al.}(1971)\citenamefont
  {Berestetskii}, \citenamefont {Lifshitz},\ and\ \citenamefont
  {Pitaevskii}}]{Rel_Quant_Theory}%
  \BibitemOpen
  \bibfield  {author} {\bibinfo {author} {\bibfnamefont {V.~B.}\ \bibnamefont
  {Berestetskii}}, \bibinfo {author} {\bibfnamefont {E.~M.}\ \bibnamefont
  {Lifshitz}}, \ and\ \bibinfo {author} {\bibfnamefont {L.~P.}\ \bibnamefont
  {Pitaevskii}},\ }\href@noop {} {\emph {\bibinfo {title} {Relativistic Quantum
  Theory}}}\ (\bibinfo  {publisher} {Pergamon Press},\ \bibinfo {year}
  {1971})\BibitemShut {NoStop}%
\bibitem [{\citenamefont {Lowdin}(1962)}]{Lowdin1962}%
  \BibitemOpen
  \bibfield  {author} {\bibinfo {author} {\bibfnamefont {P.}~\bibnamefont
  {Lowdin}},\ }\href {\doibase http://dx.doi.org/10.1063/1.1724312} {\bibfield
  {journal} {\bibinfo  {journal} {Journal of Mathematical Physics}\ }\textbf
  {\bibinfo {volume} {3}},\ \bibinfo {pages} {969} (\bibinfo {year}
  {1962})}\BibitemShut {NoStop}%
\bibitem [{\citenamefont {Krekora}\ \emph
  {et~al.}(2004{\natexlab{a}})\citenamefont {Krekora}, \citenamefont {Su},\
  and\ \citenamefont {Grobe}}]{Krekora2004}%
  \BibitemOpen
  \bibfield  {author} {\bibinfo {author} {\bibfnamefont {P.}~\bibnamefont
  {Krekora}}, \bibinfo {author} {\bibfnamefont {Q.}~\bibnamefont {Su}}, \ and\
  \bibinfo {author} {\bibfnamefont {R.}~\bibnamefont {Grobe}},\ }\href
  {\doibase 10.1103/PhysRevLett.92.040406} {\bibfield  {journal} {\bibinfo
  {journal} {Phys. Rev. Lett.}\ }\textbf {\bibinfo {volume} {92}},\ \bibinfo
  {pages} {040406} (\bibinfo {year} {2004}{\natexlab{a}})}\BibitemShut
  {NoStop}%
\bibitem [{\citenamefont {Braun}\ \emph {et~al.}(1999)\citenamefont {Braun},
  \citenamefont {Su},\ and\ \citenamefont {Grobe}}]{PhysRevA.59.604}%
  \BibitemOpen
  \bibfield  {author} {\bibinfo {author} {\bibfnamefont {J.~W.}\ \bibnamefont
  {Braun}}, \bibinfo {author} {\bibfnamefont {Q.}~\bibnamefont {Su}}, \ and\
  \bibinfo {author} {\bibfnamefont {R.}~\bibnamefont {Grobe}},\ }\href
  {\doibase 10.1103/PhysRevA.59.604} {\bibfield  {journal} {\bibinfo  {journal}
  {Phys. Rev. A}\ }\textbf {\bibinfo {volume} {59}},\ \bibinfo {pages} {604}
  (\bibinfo {year} {1999})}\BibitemShut {NoStop}%
\bibitem [{\citenamefont {Feynman}(1949)}]{Feynman1949}%
  \BibitemOpen
  \bibfield  {author} {\bibinfo {author} {\bibfnamefont {R.~P.}\ \bibnamefont
  {Feynman}},\ }\href {\doibase 10.1103/PhysRev.76.749} {\bibfield  {journal}
  {\bibinfo  {journal} {Phys. Rev.}\ }\textbf {\bibinfo {volume} {76}},\
  \bibinfo {pages} {749} (\bibinfo {year} {1949})}\BibitemShut {NoStop}%
\bibitem [{\citenamefont {Krekora}\ \emph {et~al.}(2006)\citenamefont
  {Krekora}, \citenamefont {Su},\ and\ \citenamefont
  {Grobe}}]{PhysRevA.73.022114}%
  \BibitemOpen
  \bibfield  {author} {\bibinfo {author} {\bibfnamefont {P.}~\bibnamefont
  {Krekora}}, \bibinfo {author} {\bibfnamefont {Q.}~\bibnamefont {Su}}, \ and\
  \bibinfo {author} {\bibfnamefont {R.}~\bibnamefont {Grobe}},\ }\href
  {\doibase 10.1103/PhysRevA.73.022114} {\bibfield  {journal} {\bibinfo
  {journal} {Phys. Rev. A}\ }\textbf {\bibinfo {volume} {73}},\ \bibinfo
  {pages} {022114} (\bibinfo {year} {2006})}\BibitemShut {NoStop}%
\bibitem [{\citenamefont {Castro~Neto}\ \emph {et~al.}(2009)\citenamefont
  {Castro~Neto}, \citenamefont {Guinea}, \citenamefont {Peres}, \citenamefont
  {Novoselov},\ and\ \citenamefont {Geim}}]{RevModPhys.81.109}%
  \BibitemOpen
  \bibfield  {author} {\bibinfo {author} {\bibfnamefont {A.~H.}\ \bibnamefont
  {Castro~Neto}}, \bibinfo {author} {\bibfnamefont {F.}~\bibnamefont {Guinea}},
  \bibinfo {author} {\bibfnamefont {N.~M.~R.}\ \bibnamefont {Peres}}, \bibinfo
  {author} {\bibfnamefont {K.~S.}\ \bibnamefont {Novoselov}}, \ and\ \bibinfo
  {author} {\bibfnamefont {A.~K.}\ \bibnamefont {Geim}},\ }\href {\doibase
  10.1103/RevModPhys.81.109} {\bibfield  {journal} {\bibinfo  {journal} {Rev.
  Mod. Phys.}\ }\textbf {\bibinfo {volume} {81}},\ \bibinfo {pages} {109}
  (\bibinfo {year} {2009})}\BibitemShut {NoStop}%
\bibitem [{\citenamefont {Baretty}\ \emph {et~al.}(1986)\citenamefont
  {Baretty}, \citenamefont {Ishikawa},\ and\ \citenamefont
  {Nieves}}]{Baretty1986}%
  \BibitemOpen
  \bibfield  {author} {\bibinfo {author} {\bibfnamefont {R.}~\bibnamefont
  {Baretty}}, \bibinfo {author} {\bibfnamefont {Y.}~\bibnamefont {Ishikawa}}, \
  and\ \bibinfo {author} {\bibfnamefont {J.~F.}\ \bibnamefont {Nieves}},\
  }\href {\doibase 10.1002/qua.560300713} {\bibfield  {journal} {\bibinfo
  {journal} {International Journal of Quantum Chemistry}\ }\textbf {\bibinfo
  {volume} {30}},\ \bibinfo {pages} {109} (\bibinfo {year} {1986})}\BibitemShut
  {NoStop}%
\bibitem [{\citenamefont {Gosselin}\ \emph {et~al.}(2009)\citenamefont
  {Gosselin}, \citenamefont {B{\'e}rard}, \citenamefont {Mohrbach},\ and\
  \citenamefont {Ghosh}}]{Gosselin2009}%
  \BibitemOpen
  \bibfield  {author} {\bibinfo {author} {\bibfnamefont {P.}~\bibnamefont
  {Gosselin}}, \bibinfo {author} {\bibfnamefont {A.}~\bibnamefont
  {B{\'e}rard}}, \bibinfo {author} {\bibfnamefont {H.}~\bibnamefont
  {Mohrbach}}, \ and\ \bibinfo {author} {\bibfnamefont {S.}~\bibnamefont
  {Ghosh}},\ }\href {\doibase 10.1140/epjc/s10052-008-0839-4} {\bibfield
  {journal} {\bibinfo  {journal} {The European Physical Journal C}\ }\textbf
  {\bibinfo {volume} {59}},\ \bibinfo {pages} {883} (\bibinfo {year}
  {2009})}\BibitemShut {NoStop}%
\bibitem [{\citenamefont {Bastard}(1981)}]{PhysRevB.24.5693}%
  \BibitemOpen
  \bibfield  {author} {\bibinfo {author} {\bibfnamefont {G.}~\bibnamefont
  {Bastard}},\ }\href {\doibase 10.1103/PhysRevB.24.5693} {\bibfield  {journal}
  {\bibinfo  {journal} {Phys. Rev. B}\ }\textbf {\bibinfo {volume} {24}},\
  \bibinfo {pages} {5693} (\bibinfo {year} {1981})}\BibitemShut {NoStop}%
\bibitem [{\citenamefont {Hasan}\ and\ \citenamefont
  {Kane}(2010)}]{RevModPhys.82.3045}%
  \BibitemOpen
  \bibfield  {author} {\bibinfo {author} {\bibfnamefont {M.~Z.}\ \bibnamefont
  {Hasan}}\ and\ \bibinfo {author} {\bibfnamefont {C.~L.}\ \bibnamefont
  {Kane}},\ }\href {\doibase 10.1103/RevModPhys.82.3045} {\bibfield  {journal}
  {\bibinfo  {journal} {Rev. Mod. Phys.}\ }\textbf {\bibinfo {volume} {82}},\
  \bibinfo {pages} {3045} (\bibinfo {year} {2010})}\BibitemShut {NoStop}%
\bibitem [{\citenamefont {{Deshpande}}\ and\ \citenamefont
  {{Winkler}}(2016)}]{2016arXiv160304329D}%
  \BibitemOpen
  \bibfield  {author} {\bibinfo {author} {\bibfnamefont {H.}~\bibnamefont
  {{Deshpande}}}\ and\ \bibinfo {author} {\bibfnamefont {R.}~\bibnamefont
  {{Winkler}}},\ }\href@noop {} {\bibfield  {journal} {\bibinfo  {journal}
  {ArXiv e-prints}\ } (\bibinfo {year} {2016})},\ \Eprint
  {http://arxiv.org/abs/1603.04329} {arXiv:1603.04329 [cond-mat.mes-hall]}
  \BibitemShut {NoStop}%
\bibitem [{\citenamefont {Volkov}\ and\ \citenamefont
  {Pankratov}(1985)}]{Volkov1985}%
  \BibitemOpen
  \bibfield  {author} {\bibinfo {author} {\bibfnamefont {B.~A.}\ \bibnamefont
  {Volkov}}\ and\ \bibinfo {author} {\bibfnamefont {O.~A.}\ \bibnamefont
  {Pankratov}},\ }\href@noop {} {\bibfield  {journal} {\bibinfo  {journal}
  {JETP Lett.}\ }\textbf {\bibinfo {volume} {42}},\ \bibinfo {pages} {178}
  (\bibinfo {year} {1985})}\BibitemShut {NoStop}%
\bibitem [{\citenamefont {Lee}(1981)}]{Lee1981}%
  \BibitemOpen
  \bibfield  {author} {\bibinfo {author} {\bibfnamefont {T.~D.}\ \bibnamefont
  {Lee}},\ }\href@noop {} {\emph {\bibinfo {title} {Particle physics and
  introduction to field theory}}},\ \bibinfo {edition} {2nd}\ ed.\ (\bibinfo
  {publisher} {Chur, Switzerland ; New York : Harwood Academic Publishers},\
  \bibinfo {year} {1981})\BibitemShut {NoStop}%
\bibitem [{\citenamefont {Negro}\ and\ \citenamefont
  {Tartaglia}(1981)}]{PhysRevA.23.1591}%
  \BibitemOpen
  \bibfield  {author} {\bibinfo {author} {\bibfnamefont {F.}~\bibnamefont
  {Negro}}\ and\ \bibinfo {author} {\bibfnamefont {A.}~\bibnamefont
  {Tartaglia}},\ }\href {\doibase 10.1103/PhysRevA.23.1591} {\bibfield
  {journal} {\bibinfo  {journal} {Phys. Rev. A}\ }\textbf {\bibinfo {volume}
  {23}},\ \bibinfo {pages} {1591} (\bibinfo {year} {1981})}\BibitemShut
  {NoStop}%
\bibitem [{\citenamefont {Stuckens}\ and\ \citenamefont
  {Kobe}(1986)}]{PhysRevA.34.3565}%
  \BibitemOpen
  \bibfield  {author} {\bibinfo {author} {\bibfnamefont {C.}~\bibnamefont
  {Stuckens}}\ and\ \bibinfo {author} {\bibfnamefont {D.~H.}\ \bibnamefont
  {Kobe}},\ }\href {\doibase 10.1103/PhysRevA.34.3565} {\bibfield  {journal}
  {\bibinfo  {journal} {Phys. Rev. A}\ }\textbf {\bibinfo {volume} {34}},\
  \bibinfo {pages} {3565} (\bibinfo {year} {1986})}\BibitemShut {NoStop}%
\bibitem [{\citenamefont {S\'a~Borges}\ \emph {et~al.}(1988)\citenamefont
  {S\'a~Borges}, \citenamefont {Epele}, \citenamefont {Fanchiotti},
  \citenamefont {Garc\'{\i}a~Canal},\ and\ \citenamefont
  {Simo}}]{PhysRevA.38.3101}%
  \BibitemOpen
  \bibfield  {author} {\bibinfo {author} {\bibfnamefont {J.}~\bibnamefont
  {S\'a~Borges}}, \bibinfo {author} {\bibfnamefont {L.~N.}\ \bibnamefont
  {Epele}}, \bibinfo {author} {\bibfnamefont {H.}~\bibnamefont {Fanchiotti}},
  \bibinfo {author} {\bibfnamefont {C.~A.}\ \bibnamefont {Garc\'{\i}a~Canal}},
  \ and\ \bibinfo {author} {\bibfnamefont {F.~R.~A.}\ \bibnamefont {Simo}},\
  }\href {\doibase 10.1103/PhysRevA.38.3101} {\bibfield  {journal} {\bibinfo
  {journal} {Phys. Rev. A}\ }\textbf {\bibinfo {volume} {38}},\ \bibinfo
  {pages} {3101} (\bibinfo {year} {1988})}\BibitemShut {NoStop}%
\bibitem [{\citenamefont {Chen}\ and\ \citenamefont
  {Chiou}(2014{\natexlab{b}})}]{PhysRevA.89.032111}%
  \BibitemOpen
  \bibfield  {author} {\bibinfo {author} {\bibfnamefont {T.-W.}\ \bibnamefont
  {Chen}}\ and\ \bibinfo {author} {\bibfnamefont {D.-W.}\ \bibnamefont
  {Chiou}},\ }\href {\doibase 10.1103/PhysRevA.89.032111} {\bibfield  {journal}
  {\bibinfo  {journal} {Phys. Rev. A}\ }\textbf {\bibinfo {volume} {89}},\
  \bibinfo {pages} {032111} (\bibinfo {year} {2014}{\natexlab{b}})}\BibitemShut
  {NoStop}%
\bibitem [{\citenamefont {Reiher}(2006)}]{Reiher2006}%
  \BibitemOpen
  \bibfield  {author} {\bibinfo {author} {\bibfnamefont {M.}~\bibnamefont
  {Reiher}},\ }\href {\doibase 10.1007/s00214-005-0003-2} {\bibfield  {journal}
  {\bibinfo  {journal} {Theoretical Chemistry Accounts}\ }\textbf {\bibinfo
  {volume} {116}},\ \bibinfo {pages} {241} (\bibinfo {year}
  {2006})}\BibitemShut {NoStop}%
\bibitem [{\citenamefont {Winkler}\ and\ \citenamefont
  {R\"ossler}(1993)}]{Winkler1993}%
  \BibitemOpen
  \bibfield  {author} {\bibinfo {author} {\bibfnamefont {R.}~\bibnamefont
  {Winkler}}\ and\ \bibinfo {author} {\bibfnamefont {U.}~\bibnamefont
  {R\"ossler}},\ }\href {\doibase 10.1103/PhysRevB.48.8918} {\bibfield
  {journal} {\bibinfo  {journal} {Phys. Rev. B}\ }\textbf {\bibinfo {volume}
  {48}},\ \bibinfo {pages} {8918} (\bibinfo {year} {1993})}\BibitemShut
  {NoStop}%
\bibitem [{\citenamefont {Shon}\ and\ \citenamefont {Ando}(1998)}]{Ando1998}%
  \BibitemOpen
  \bibfield  {author} {\bibinfo {author} {\bibfnamefont {N.}~\bibnamefont
  {Shon}}\ and\ \bibinfo {author} {\bibfnamefont {T.}~\bibnamefont {Ando}},\
  }\href {\doibase 10.1143/JPSJ.67.2421} {\bibfield  {journal} {\bibinfo
  {journal} {J. Phys. Soc. Jpn.}\ }\textbf {\bibinfo {volume} {67}},\ \bibinfo
  {pages} {2421} (\bibinfo {year} {1998})}\BibitemShut {NoStop}%
\bibitem [{\citenamefont {Yao}\ \emph {et~al.}(2009)\citenamefont {Yao},
  \citenamefont {Yang},\ and\ \citenamefont {Niu}}]{PhysRevLett.102.096801}%
  \BibitemOpen
  \bibfield  {author} {\bibinfo {author} {\bibfnamefont {W.}~\bibnamefont
  {Yao}}, \bibinfo {author} {\bibfnamefont {S.~A.}\ \bibnamefont {Yang}}, \
  and\ \bibinfo {author} {\bibfnamefont {Q.}~\bibnamefont {Niu}},\ }\href
  {\doibase 10.1103/PhysRevLett.102.096801} {\bibfield  {journal} {\bibinfo
  {journal} {Phys. Rev. Lett.}\ }\textbf {\bibinfo {volume} {102}},\ \bibinfo
  {pages} {096801} (\bibinfo {year} {2009})}\BibitemShut {NoStop}%
\bibitem [{\citenamefont {Andreev}(1964)}]{Andreev1964}%
  \BibitemOpen
  \bibfield  {author} {\bibinfo {author} {\bibfnamefont {A.~F.}\ \bibnamefont
  {Andreev}},\ }\href@noop {} {\bibfield  {journal} {\bibinfo  {journal} {Sov.
  Phys. JETP}\ }\textbf {\bibinfo {volume} {19}},\ \bibinfo {pages} {1228}
  (\bibinfo {year} {1964})}\BibitemShut {NoStop}%
\bibitem [{\citenamefont {Zhang}\ \emph {et~al.}(2009)\citenamefont {Zhang},
  \citenamefont {Brar}, \citenamefont {Girit}, \citenamefont {Zettl},\ and\
  \citenamefont {Crommie}}]{Zhang2009}%
  \BibitemOpen
  \bibfield  {author} {\bibinfo {author} {\bibfnamefont {Y.}~\bibnamefont
  {Zhang}}, \bibinfo {author} {\bibfnamefont {V.}~\bibnamefont {Brar}},
  \bibinfo {author} {\bibfnamefont {C.}~\bibnamefont {Girit}}, \bibinfo
  {author} {\bibfnamefont {A.}~\bibnamefont {Zettl}}, \ and\ \bibinfo {author}
  {\bibfnamefont {M.}~\bibnamefont {Crommie}},\ }\href@noop {} {\bibfield
  {journal} {\bibinfo  {journal} {Nat. Phys.}\ }\textbf {\bibinfo {volume}
  {5}},\ \bibinfo {pages} {722} (\bibinfo {year} {2009})}\BibitemShut {NoStop}%
\bibitem [{\citenamefont {Bychkov}\ and\ \citenamefont
  {Rashba}(1984)}]{Rashba1984}%
  \BibitemOpen
  \bibfield  {author} {\bibinfo {author} {\bibfnamefont {Y.~A.}\ \bibnamefont
  {Bychkov}}\ and\ \bibinfo {author} {\bibfnamefont {E.~I.}\ \bibnamefont
  {Rashba}},\ }\href@noop {} {\bibfield  {journal} {\bibinfo  {journal} {J.
  Phys. C}\ }\textbf {\bibinfo {volume} {17}},\ \bibinfo {pages} {6039}
  (\bibinfo {year} {1984})}\BibitemShut {NoStop}%
\bibitem [{\citenamefont {{Katsnelson, M. I.}}(2006)}]{Katsnelson2006}%
  \BibitemOpen
  \bibfield  {author} {\bibinfo {author} {\bibnamefont {{Katsnelson, M. I.}}},\
  }\href {\doibase 10.1140/epjb/e2006-00203-1} {\bibfield  {journal} {\bibinfo
  {journal} {Eur. Phys. J. B}\ }\textbf {\bibinfo {volume} {51}},\ \bibinfo
  {pages} {157} (\bibinfo {year} {2006})}\BibitemShut {NoStop}%
\bibitem [{\citenamefont {Krekora}\ \emph {et~al.}(2005)\citenamefont
  {Krekora}, \citenamefont {Su},\ and\ \citenamefont
  {Grobe}}]{PhysRevA.72.064103}%
  \BibitemOpen
  \bibfield  {author} {\bibinfo {author} {\bibfnamefont {P.}~\bibnamefont
  {Krekora}}, \bibinfo {author} {\bibfnamefont {Q.}~\bibnamefont {Su}}, \ and\
  \bibinfo {author} {\bibfnamefont {R.}~\bibnamefont {Grobe}},\ }\href
  {\doibase 10.1103/PhysRevA.72.064103} {\bibfield  {journal} {\bibinfo
  {journal} {Phys. Rev. A}\ }\textbf {\bibinfo {volume} {72}},\ \bibinfo
  {pages} {064103} (\bibinfo {year} {2005})}\BibitemShut {NoStop}%
\bibitem [{\citenamefont {Zak}(1989)}]{PhysRevLett.62.2747}%
  \BibitemOpen
  \bibfield  {author} {\bibinfo {author} {\bibfnamefont {J.}~\bibnamefont
  {Zak}},\ }\href {\doibase 10.1103/PhysRevLett.62.2747} {\bibfield  {journal}
  {\bibinfo  {journal} {Phys. Rev. Lett.}\ }\textbf {\bibinfo {volume} {62}},\
  \bibinfo {pages} {2747} (\bibinfo {year} {1989})}\BibitemShut {NoStop}%
\bibitem [{\citenamefont {B\'erard}\ and\ \citenamefont
  {Mohrbach}(2004)}]{PhysRevD.69.127701}%
  \BibitemOpen
  \bibfield  {author} {\bibinfo {author} {\bibfnamefont {A.}~\bibnamefont
  {B\'erard}}\ and\ \bibinfo {author} {\bibfnamefont {H.}~\bibnamefont
  {Mohrbach}},\ }\href {\doibase 10.1103/PhysRevD.69.127701} {\bibfield
  {journal} {\bibinfo  {journal} {Phys. Rev. D}\ }\textbf {\bibinfo {volume}
  {69}},\ \bibinfo {pages} {127701} (\bibinfo {year} {2004})}\BibitemShut
  {NoStop}%
\bibitem [{\citenamefont {Bacry}(1988)}]{Bacry1988}%
  \BibitemOpen
  \bibfield  {author} {\bibinfo {author} {\bibfnamefont {H.}~\bibnamefont
  {Bacry}},\ }\href@noop {} {\emph {\bibinfo {title} {”Localizability and
  Space in Quantum Physics}}},\ Vol.\ \bibinfo {volume} {308}\ (\bibinfo
  {publisher} {Heidelberg, Springer-Verlag},\ \bibinfo {year}
  {1988})\BibitemShut {NoStop}%
\bibitem [{\citenamefont {Bérard}\ and\ \citenamefont
  {Mohrbach}(2006)}]{Berard2006190}%
  \BibitemOpen
  \bibfield  {author} {\bibinfo {author} {\bibfnamefont {A.}~\bibnamefont
  {Bérard}}\ and\ \bibinfo {author} {\bibfnamefont {H.}~\bibnamefont
  {Mohrbach}},\ }\href {\doibase
  http://dx.doi.org/10.1016/j.physleta.2005.11.071} {\bibfield  {journal}
  {\bibinfo  {journal} {Physics Letters A}\ }\textbf {\bibinfo {volume}
  {352}},\ \bibinfo {pages} {190 } (\bibinfo {year} {2006})}\BibitemShut
  {NoStop}%
\bibitem [{\citenamefont {D\'avid}\ and\ \citenamefont
  {Cserti}(2010)}]{PhysRevB.81.121417}%
  \BibitemOpen
  \bibfield  {author} {\bibinfo {author} {\bibfnamefont {G.}~\bibnamefont
  {D\'avid}}\ and\ \bibinfo {author} {\bibfnamefont {J.}~\bibnamefont
  {Cserti}},\ }\href {\doibase 10.1103/PhysRevB.81.121417} {\bibfield
  {journal} {\bibinfo  {journal} {Phys. Rev. B}\ }\textbf {\bibinfo {volume}
  {81}},\ \bibinfo {pages} {121417} (\bibinfo {year} {2010})}\BibitemShut
  {NoStop}%
\bibitem [{\citenamefont {Krekora}\ \emph
  {et~al.}(2004{\natexlab{b}})\citenamefont {Krekora}, \citenamefont {Su},\
  and\ \citenamefont {Grobe}}]{PhysRevA.70.054101}%
  \BibitemOpen
  \bibfield  {author} {\bibinfo {author} {\bibfnamefont {P.}~\bibnamefont
  {Krekora}}, \bibinfo {author} {\bibfnamefont {Q.}~\bibnamefont {Su}}, \ and\
  \bibinfo {author} {\bibfnamefont {R.}~\bibnamefont {Grobe}},\ }\href
  {\doibase 10.1103/PhysRevA.70.054101} {\bibfield  {journal} {\bibinfo
  {journal} {Phys. Rev. A}\ }\textbf {\bibinfo {volume} {70}},\ \bibinfo
  {pages} {054101} (\bibinfo {year} {2004}{\natexlab{b}})}\BibitemShut
  {NoStop}%
\bibitem [{\citenamefont {Schweber}(1962)}]{Schweber1962}%
  \BibitemOpen
  \bibfield  {author} {\bibinfo {author} {\bibfnamefont {S.}~\bibnamefont
  {Schweber}},\ }\href@noop {} {\emph {\bibinfo {title} {An Introduction to
  Relativistic Quantum Field theory}}}\ (\bibinfo  {publisher} {Harper and
  Row},\ \bibinfo {address} {New York},\ \bibinfo {year} {1962})\BibitemShut
  {NoStop}%
\bibitem [{\citenamefont {Holstein}(1998)}]{Holstein1998}%
  \BibitemOpen
  \bibfield  {author} {\bibinfo {author} {\bibfnamefont {B.~R.}\ \bibnamefont
  {Holstein}},\ }\href@noop {} {\bibfield  {journal} {\bibinfo  {journal}
  {American Journal of Physics}\ }\textbf {\bibinfo {volume} {66}} (\bibinfo
  {year} {1998})}\BibitemShut {NoStop}%
\bibitem [{\citenamefont {Urru}\ \emph {et~al.}(2015)\citenamefont {Urru},
  \citenamefont {Cocco},\ and\ \citenamefont {Fiorentini}}]{Urru2015}%
  \BibitemOpen
  \bibfield  {author} {\bibinfo {author} {\bibfnamefont {A.}~\bibnamefont
  {Urru}}, \bibinfo {author} {\bibfnamefont {G.}~\bibnamefont {Cocco}}, \ and\
  \bibinfo {author} {\bibfnamefont {V.}~\bibnamefont {Fiorentini}},\
  }\href@noop {} {\bibfield  {journal} {\bibinfo  {journal}
  {arXiv:1511.01341v1}\ } (\bibinfo {year} {2015})}\BibitemShut {NoStop}%
\bibitem [{\citenamefont {Newton}\ and\ \citenamefont
  {Wigner}(1949)}]{RevModPhys.21.400}%
  \BibitemOpen
  \bibfield  {author} {\bibinfo {author} {\bibfnamefont {T.~D.}\ \bibnamefont
  {Newton}}\ and\ \bibinfo {author} {\bibfnamefont {E.~P.}\ \bibnamefont
  {Wigner}},\ }\href {\doibase 10.1103/RevModPhys.21.400} {\bibfield  {journal}
  {\bibinfo  {journal} {Rev. Mod. Phys.}\ }\textbf {\bibinfo {volume} {21}},\
  \bibinfo {pages} {400} (\bibinfo {year} {1949})}\BibitemShut {NoStop}%
\bibitem [{\citenamefont {Zawadzki}(2005)}]{PhysRevB.72.085217}%
  \BibitemOpen
  \bibfield  {author} {\bibinfo {author} {\bibfnamefont {W.}~\bibnamefont
  {Zawadzki}},\ }\href {\doibase 10.1103/PhysRevB.72.085217} {\bibfield
  {journal} {\bibinfo  {journal} {Phys. Rev. B}\ }\textbf {\bibinfo {volume}
  {72}},\ \bibinfo {pages} {085217} (\bibinfo {year} {2005})}\BibitemShut
  {NoStop}%
\bibitem [{\citenamefont {Rusin}\ and\ \citenamefont
  {Zawadzki}(2012{\natexlab{a}})}]{Zawadzki2012}%
  \BibitemOpen
  \bibfield  {author} {\bibinfo {author} {\bibfnamefont {T.~M.}\ \bibnamefont
  {Rusin}}\ and\ \bibinfo {author} {\bibfnamefont {W.}~\bibnamefont
  {Zawadzki}},\ }\href@noop {} {\bibfield  {journal} {\bibinfo  {journal}
  {Journal of Physics A: Mathematical and Theoretical}\ }\textbf {\bibinfo
  {volume} {45}},\ \bibinfo {pages} {315301} (\bibinfo {year}
  {2012}{\natexlab{a}})}\BibitemShut {NoStop}%
\bibitem [{\citenamefont {Hardekopf}\ and\ \citenamefont
  {Sucher}(1984)}]{PhysRevA.30.703}%
  \BibitemOpen
  \bibfield  {author} {\bibinfo {author} {\bibfnamefont {G.}~\bibnamefont
  {Hardekopf}}\ and\ \bibinfo {author} {\bibfnamefont {J.}~\bibnamefont
  {Sucher}},\ }\href {\doibase 10.1103/PhysRevA.30.703} {\bibfield  {journal}
  {\bibinfo  {journal} {Phys. Rev. A}\ }\textbf {\bibinfo {volume} {30}},\
  \bibinfo {pages} {703} (\bibinfo {year} {1984})}\BibitemShut {NoStop}%
\bibitem [{\citenamefont {Neznamov}\ and\ \citenamefont
  {Silenko}(2009)}]{Neznamov2009}%
  \BibitemOpen
  \bibfield  {author} {\bibinfo {author} {\bibfnamefont {V.~P.}\ \bibnamefont
  {Neznamov}}\ and\ \bibinfo {author} {\bibfnamefont {A.~J.}\ \bibnamefont
  {Silenko}},\ }\href {\doibase http://dx.doi.org/10.1063/1.3268592} {\bibfield
   {journal} {\bibinfo  {journal} {Journal of Mathematical Physics}\ }\textbf
  {\bibinfo {volume} {50}},\ \bibinfo {eid} {122302} (\bibinfo {year} {2009}),\
  http://dx.doi.org/10.1063/1.3268592}\BibitemShut {NoStop}%
\bibitem [{\citenamefont {Rusin}\ and\ \citenamefont
  {Zawadzki}(2012{\natexlab{b}})}]{PhysRevA.86.032103}%
  \BibitemOpen
  \bibfield  {author} {\bibinfo {author} {\bibfnamefont {T.~M.}\ \bibnamefont
  {Rusin}}\ and\ \bibinfo {author} {\bibfnamefont {W.}~\bibnamefont
  {Zawadzki}},\ }\href {\doibase 10.1103/PhysRevA.86.032103} {\bibfield
  {journal} {\bibinfo  {journal} {Phys. Rev. A}\ }\textbf {\bibinfo {volume}
  {86}},\ \bibinfo {pages} {032103} (\bibinfo {year}
  {2012}{\natexlab{b}})}\BibitemShut {NoStop}%
\end{thebibliography}
\end{document}